\documentclass{aa}  

\usepackage{graphicx}
\usepackage{txfonts}
\usepackage{xcolor,color}
\usepackage{comment}

\newcommand{\otrain}{O'TRAIN}
\begin{document}

   \title{\otrain : a robust and flexible Real/Bogus classifier for the study of the optical transient sky }
   \titlerunning{O'TRAIN, for the study of the optical transient sky}
   \subtitle{}

   \author{K. Makhlouf\inst{1,2},
          D. Turpin\inst{1},
          D. Corre\inst{3},
          S. Karpov\inst{4},
          D. A. Kann\inst{5}
          \and
          A. Klotz\inst{6,7}
          }
    \authorrunning{K. Makhlouf et al.}
   \institute{Université Paris-Saclay, CNRS, CEA, Astrophysique, Instrumentation et Modélisation de Paris-Saclay, 91191, Gif-sur-Yvette, France\\
              \email{damien.turpin@cea.fr}
         \and
            Ecole Centrale Lille, Cité scientifique CS 20048, 59651 Villeneuve d'Ascq cedex, France
         \and
             Sorbonne Université, CNRS, UMR 7095, Institut d’Astrophysique de Paris, 98 bis boulevard Arago, 75014 Paris, France
        \and
            CEICO, Institute of Physics of  the  Czech  Academy  of Sciences, Prague, Czech Republic
        \and
            Instituto de Astrof\'isica de Andaluc\'ia (IAA-CSIC), Glorieta de la Astronom\'ia s/n, 18008 Granada, Spain
        \and
            IRAP, Université de Toulouse, CNRS, UPS, 14 Avenue Edouard Belin, F-31400 Toulouse, France
        \and
            Université Paul Sabatier Toulouse III, Université de Toulouse, 118 route de Narbonne, 31400 Toulouse, France
             }

   \date{Received December 19, 2021; accepted XX, 2022}

 
  \abstract
   {The scientific interest in studying high-energy transient phenomena in the Universe has largely grown for the last decade. Now, multiple ground-based survey projects have emerged to continuously monitor the optical (and multi-messenger) transient sky at higher image cadences and cover always larger portions of the sky every night. These novel approaches lead to a huge increase of the global alert rates which need to be handled with care especially by keeping the false alarms as low as possible. Therefore, the standard transient detection pipelines previously designed for narrow field of view instruments must now integrate more sophisticated tools to deal with the growing number and diversity of alerts and false alarms.}
   {Deep machine learning algorithms have now proven their efficiency in recognizing patterns in images. These methods are now used in astrophysics to perform different classification tasks such as identifying bogus from real transient point like sources. We explore this method to provide a robust and flexible algorithm that could be included in any kind of transient detection pipeline.}
   {We built a Convolutional Neural Network (CNN) algorithm in order to perform a real/bogus classification task on transient candidate cutouts (subtraction residuals) provided by different kinds of optical telescopes. The training involved human-supervised labeling of the cutouts, which had been split in two balanced data sets with \textit{True} and \textit{False} point-like source candidates. We tested our CNN model 
   on the candidates produced by two different transient detection pipelines.
   In addition, we provide several diagnostic tools to evaluate the classification performance of our CNN models.}
   {We show that our CNN algorithm can be successfully trained on a large diversity of images having very different pixel scales. In this training process, we did not detect any strong over(under)fitting with the requirement of providing cutouts with a limited size no larger than $50\times50$ pixels. Tested on optical images from four different telescopes and utilising two different transient detection pipelines, our CNN model provides robust real/bogus classification performance accuracy from 93\% up to 98\% of well classified candidates.}
   {}


   \maketitle
%

\section{Introduction}

Time-domain astronomy consists of studying variable sources in the Universe that periodically change in brightness or transient sources coming from the sudden outburst of known flaring objects or single cataclysmic events. For the past few years, our way of observing and monitoring the optical transient sky has significantly evolved both with the arrival of new optical synoptic survey projects and the birth of multi-messenger astronomy. It is now a matter of observing increasingly larger portions of the sky at higher sensitivities with a high image cadence. New optical synoptic surveys such as ATLAS \citep{Tonry18}, GOTO \citep{Dyer20}, MeerLICHT and BlackGEM \citep{Groot19}, SVOM/GWAC \citep{Han21}, ZTF \citep{Bellm19,Graham2018} or the upcoming Vera Rubin/LSST \citep{Ivezic19} are now able to observe their entire observable sky in very few nights. Thanks to their large fields of view (FoV) and their high image cadences at moderate and high sensitivities, a fraction of their observation time can also be dedicated to the follow-up of poorly localized multi-messenger alerts sent by gravitational wave detectors or high-energy neutrino and gamma-ray telescopes. The current optical surveys already lead to the discovery of tens of thousands of new optical transients. Those transients in part belong to known astrophysical classes such as supernovae or galactic flaring stars while new classes of transients have been recently discovered. For example, since 2018, the ATLAS and ZTF surveys have already confirmed the existence of a new type of transient called \textit{Fast Blue Optical Transients, (FBOTs)} for which only 4 events have been robustly identified so far: ATcow18 \citep{Smartt18,Prentice18}, ZTF18abvkwla \citep{Ho20}, CSS161010 \citep{Coppejans20} and AT2020xnd/ZTF20acigmel \citep{Bright21,Perley21,Ho21}. Down to the minute timescales, the SVOM/GWAC fast cadence survey \citep{Han21} has also detected new powerful outburts from nearby M dwarf stars \citep{Xin21,Wang21}. Finally, a new class of luminous supernova explosion, the so-called super luminous supernovae (SLSNe), have also been identified only a decade ago with currently about 100 candidates reported and around 20 being extensively studied \citep{Quimby11,GalYam12,GalYam19}. Actually, we are just starting to extensively explore the zoo of optical transients especially towards the shortest timescales (minute to hour timescales). Hence, there is room for new discoveries by adapting the different survey or follow-up observational strategies (image cadence, filter strategy, frequency of revisits, limiting magnitude depth, etc.) as well as the algorithm used for identifying the different types of transients.
Behind the scene, there are actually an unprecedented amount of images and data to process as well as transient candidates to characterise every night. As an example, a survey like ZTF can typically produce up to $2\cdot 10^6$ raw alerts at 5$\sigma$ confidence level every clear night, among which $10^3-10^5$ are likely real sources \citep{Masci19} . This is nothing in comparison to the millions of real alerts (corresponding to 20 terabytes of data every night) the Vera Rubin LSST survey will produce daily during its ten years of operations. As a consequence, new detection pipelines, used for detecting and identifying in near real-time new sources in optical images, have to be designed by taking in consideration these new alert flow constraints. To do so, many of them now make use of (deep) machine learning (ML) methods inspired from big data problems. Deep ML algorithms allow to quickly process astrophysical images or different sets of complex informations to give classification probabilities in near real-time. These probabilities can then be easily interpreted by astronomers to trigger further targeted follow-up observations. Usually, in time-domain astronomy, ML algorithms are used to perform the two following classification tasks:
\begin{enumerate}
    \item \textbf{Real / bogus classification} to reduce the false alarm rate due to artefacts that may be falsely identified as new real sources \citep{Gieseke17,Duev19,Turpin20,Killestein21,Hosenie21}
    
    \item \textbf{Astrophysical classification} to identify the types of transients the detection pipelines have detected based on several pieces of information about some key parameters evolving with time, the flux-color evolution of the transient or its spectral shape and associated features, etc. \citep{CarrascoDavis19,Moller20,CarrascoDavis20,Burhanudin21}
\end{enumerate}
These fast classifiers are now a standard in the software architecture of any detection pipeline to avoid the need for humans in the loop as much as possible. This is particularly relevant when is is about performing systematic manual visual inspections of the transient candidate cutouts as it is a heavily time-consuming and sometimes complex task. In addition, when talking about processing big data sets, the ML algorithms generally turn out to be more reliable and stable than human decision-taking yielding better scientific perspectives for each promising transient candidate.

In this paper, we propose a robust machine learning algorithm called \otrain\, for Optical TRAnsient Identification Network, to filter out any type of bogus from a list of optical transient (OT) candidates a detection pipeline may output. Our real/bogus (RB) classifier is based on a Convolutional Neural Network (CNN) algorithm, a method that already proved its efficiency in such a classification task \citep{Gieseke17, Turpin20}. We developed a lot of pedagogical tools to easily launch a training procedure and diagnose the classification performances. Therefore, we provide a generic and flexible classifier that can be embedded in any detection pipeline and applied to a broad range of image characteristics. In section \ref{sec:det_pipe}, we will briefly describe the general task a detection pipeline is supposed to perform and the expected outputs that will be used by our CNN model. We will then detail, in section \ref{sec:otrain}, the architecture of our CNN and the implementation setup to perform the training in section \ref{sec:training}. The different data sets and the diagnosis tools used to evaluate the performances of our CNN architecture are described in sections \ref{sec:practical_cases} and \ref{sec:tools}, respectively. We show our results in section \ref{sec:results} and we finally discuss some important future prospects for our work in section \ref{sec:discussion}. Our conclusions are given in section \ref{sec:conclusion}.

\section{Transient detection pipeline : inputs and outputs}
\label{sec:det_pipe}
To find new transient sources in optical images, a detection pipeline usually acts on the so-called \textit{science-ready} images or pre-processed images. These images are actually the raw images corrected from the dark, bias, offset and flat field  master images and for which the astrometry has been calibrated with the WCS coordinates system using dedicated softwares like \textit{SCAMP}\footnote{\url{https://www.astromatic.net/software/scamp/}} \citep{Bertin06} or Astrometry.net\footnote{\url{https://astrometry.net/}}. A second pre-processing step is usually needed to obtain a photometric calibration of the scientific images. This is done in several steps:
\begin{enumerate}
    \item Extract the position of the point-like sources using, for example, the \textit{SExtractor} software \citep{Bertin96} or the python library SEP \citep{Barbary16}.
    \item Estimate the Point-Spread Function (PSF) model of the image or in different regions of the images (PSF varying model) using softwares like \textit{PSFEx} \citep{Bertin11,psfex}.
    \item Subtract the image background and mask the brightest (saturated) stars that can lead to errors when measuring both their centroid positions and total flux in Analog Digital Units (ADU). This tasks can make use of the \textit{Photutils}\footnote{\url{https://github.com/astropy/photutils}} Astropy package \citep{Bradley21} and \textit{PSFEx}.
    \item Cross-match the detected sources in the image with known star catalogues and build the photometric model for their instrumental magnitudes. This is usually done using the Xmatch \citep{Boch12,Pineau20} and Vizier services of the CDS, Strasbourg, France. 
\end{enumerate}

After these steps, the science-ready images can be exploited to search for new transient sources. Some detection pipelines might also add few other steps by already removing most of the cosmic-ray induced-artefacts or identifying moving objects (asteroids and solar system objects). Two methods are traditionally employed to find OT candidates : the catalog cross-matching and the difference image analysis.
\subsection{The catalog cross matching method}
This method consists in searching for uncatalogued sources in the scientific images. The list of sources and their positions extracted by \textit{SExtractor} are compared with the positions of known stars by associating both sources within a given cross-match radius (typically few arcseconds). To be able to compare all the extracted sources with a given catalog, one needs to choose the catalogs that have deeper image sensitivities than the science images. This method is easy to set up and to apply to a large amount of images but suffers from two main important limitations. First, depending on the image pixel scale and the accuracy of the astrometric calibration of the science images, this method may hardly distinguished blended sources leading either to false positive cross matchings or wrong mismatches. Secondly, for flaring or variable unknown sources, this method is limited to the detection of only large flux amplitude variations between the science and the reference images. For these reasons, the catalog cross matching method would generally yield an incomplete list of transient candidates. This list must be completed by a difference image analysis to collect the transient sources close to catalogued sources and/or having faint variations in brightness.

\subsection{The difference image analysis}
By definition, optical transients are new sources that suddenly imprint their patterns in the images with their fluxes strongly evolving with time. Performing the subtraction of the science images with some reference images allows to reveal the presence of these new sources as significant excesses in the residual images. While being a powerful tool to identify transient and flux-varying sources, it implies to perform several important steps, which require fine tuning, prior to the subtraction. The reference images have to be carefully chosen in order to not contain the transient sources. Depending on the timescale of the OT flux evolution, the reference images are usually taken days to months prior to the science images. A reference image can originate either from the same telescope which acquired the science image or from all-sky surveys providing public image databases such as PanSTARRS \citep{Chambers16} or 2MASS \citep{Skrutskie06}. These catalogs of reference images can be directly downloaded from a catalog server or extracted by using the hips2fits\footnote{\url{https://alasky.u-strasbg.fr/hips-image-services/hips2fits}} service at the CDS, Strasbourg. For such a subtraction technique, it is preferable to obtain reference images of deeper limiting magnitudes and better seeing than the science images as well as taken with the same or at least close filter system. A bad pixel map for the reference image can also be produced to already remove bad pixel values or saturated stars in the subtraction process. The science and reference images must be well aligned and the PSF resampled if the reference images originate from an all-sky survey having different pixel scales. A flux normalization of the two sets of images are also performed to ensure the good quality of residual images. These steps and the image subtraction can be done by softwares such as \textit{Montage}\footnote{\url{http://montage.ipac.caltech.edu/docs/montagePy-UG.html}} (image realignment) and \textit{HOTPANTS}\footnote{\url{https://github.com/acbecker/hotpants}}\citep{hotpants}. 

\subsection{Minimum output required for a transient detection pipeline}

Both of the methods mentioned above will output two lists of OT candidates that will then be merged into a single one to avoid redundancies. This final list basically describes the properties of each OT candidate such as:
\begin{itemize}
\item their celestial and physical coordinates in the image
\item their Full Width at Half Maximum or FWHM
\item their measured magnitude with the associated error
\item their detection time
\item additional information and flags that may help to classify them (edge position and extended object flags, Signal-to-Noise ratio or SNR, light curve, etc.)
\end{itemize}
In addition, small (typically few tens of pixels) cutouts centered at the position of the OT candidates are usually cropped from the original, reference and residual images for a manual visual inspection by a scientific expert. After the visual inspection, the OT candidates are kept either as promising sources to be followed-up or discarded. This selection task is exactly what a RB classifier must efficiently do. These cutouts and primarily the residuals will be the input for our RB classifier to deliver its decision.

\subsection{Sources of artefacts}

The various artefacts one can find in astronomical images are usually produced during three main steps. First, during the acquisition, the images can be contaminated by cosmic ray tracks, blooming or crosstalk effects from the saturated and bright stars, artificial sources or tracks left by human-made flying objects (satellites, planes, etc.), hot or bad groups of pixels. The so called "Ghost" sources (diffuse extended sources) are also observed in optical images due to multiple light reflections from the optical system back to the CCD chip. Secondly, at the CCD reading step, some issues can lead to some columns of pixels that are no longer exploitable. Finally, some artefacts can appear when subtracting two images either due to bad image alignment or a lack of optimization of the \textit{HOTPANTS} parameters especially for the subtraction of survey catalog images which have a different pixel scales compared to the science image. At the end, all these processes create a very large and diverse collection of artefacts that may lead to many false detections. While some artefacts are easily recognizable, others can hardly be distinguished from point-like sources by eye or with standard filtering systems such as PSF-matching methods.
\section{\otrain : a Real/Bogus CNN classifer}
\label{sec:otrain}
\subsection{Purpose of \otrain}
Artificial intelligence models are built to lighten the workload of the astronomers on duty and to help them with the decision making. They take on annotated data and try to understand the logic behind them by creating connections between their properties. Our work involves cutouts centered at the position of the OT candidates that we want to classify into two folders: real transients and bogus. The most dominant model in this sort of computer vision tasks is the convolutional neural network since they use every information in the input image (the pixels) without being computationally expensive. By applying filters, The CNN highlights regions of the image according to their relevance when classifying the image, hence the name convolutional \citep{Yamashita18}. These regions are commonly called "features". They would distinguish, in our case, the real sources from the bogus.\\In this section, we will go into more details about our model architecture, its configuration settings and how it learns to decide the relevant features of the source, and their spatial hierarchies.
\subsection{Model architecture}
\begin{figure*}[h!]
    \centering
    \includegraphics[width=0.7\textwidth]{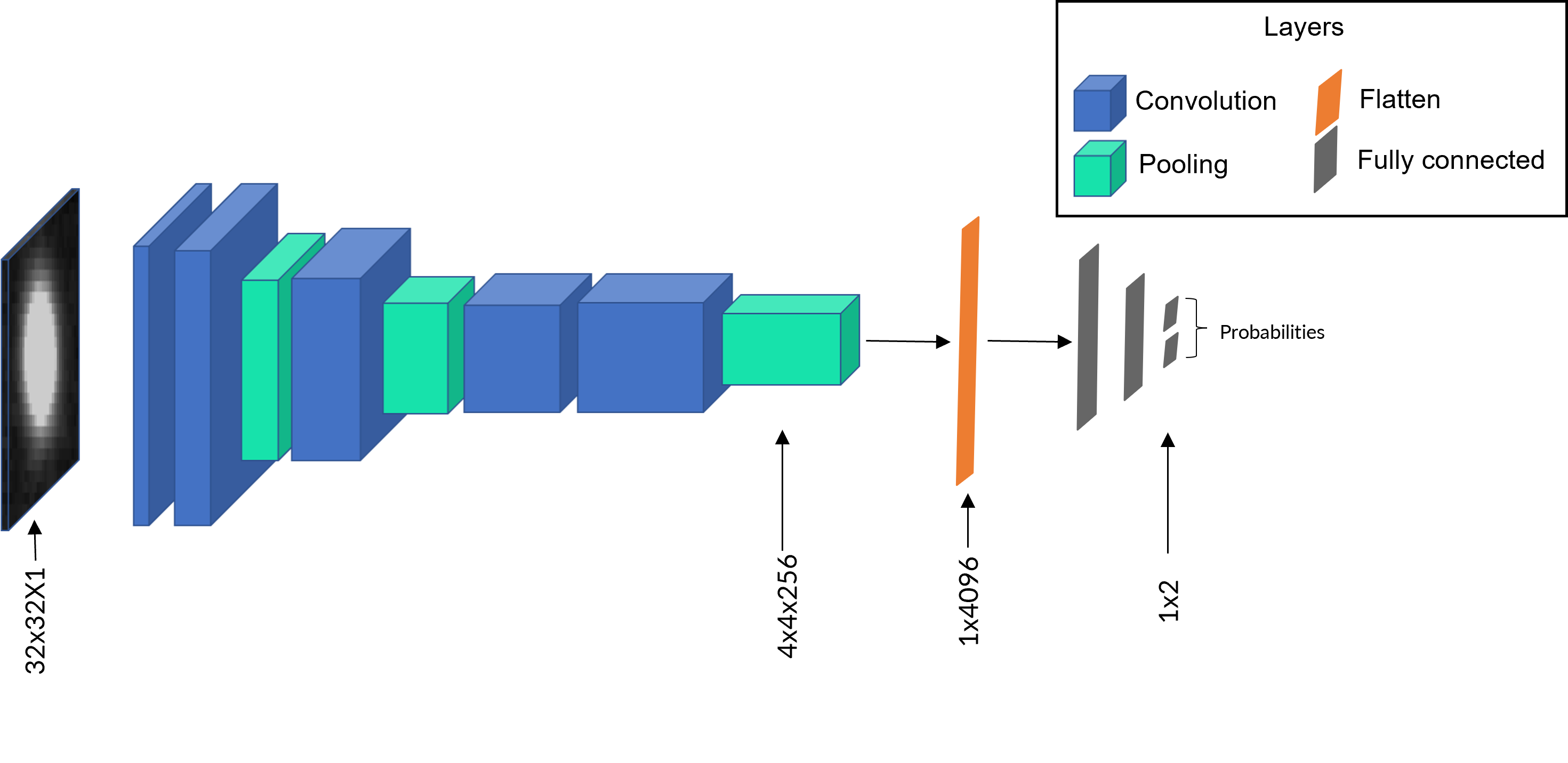}
    \caption{Illustration of the \otrain\ model architecture for performing the binary classification task : real transients or bogus.}
    \label{fig:CNN_schema}
\end{figure*}

We start off by extracting the features. At each convolutional layer, multiple filters are applied to the input of the layer, the results of these convolutions are called a feature maps. This output goes through a pooling layer, that enhances the features, while maintaining the spatial information. The final output feature maps are then fed to fully connected layers (i.e., Dense layers) thus connecting the features to the corresponding class. The number of filters in each convolutional layer, their sizes, the number of cells in the dense layers, are all configurations that should be decided to optimize the model's performance on our task. Since our cutouts are small (roughly few tens of pixels), we chose to apply 3$\times$3 filters in our convolutional layers. The number of filters is increasing throughout the network. The more filters we use the more features we extract. As we do not want to extract features from the noise next to the source, we limit the number of filters in the first layers. In Figure \ref{fig:CNN_schema}, we illustrate the different layers of our CNN in which the input images will be analyzed step by step.

\subsection{Training procedure}

\begin{figure}
    \centering
    \includegraphics[width=1.0\columnwidth]{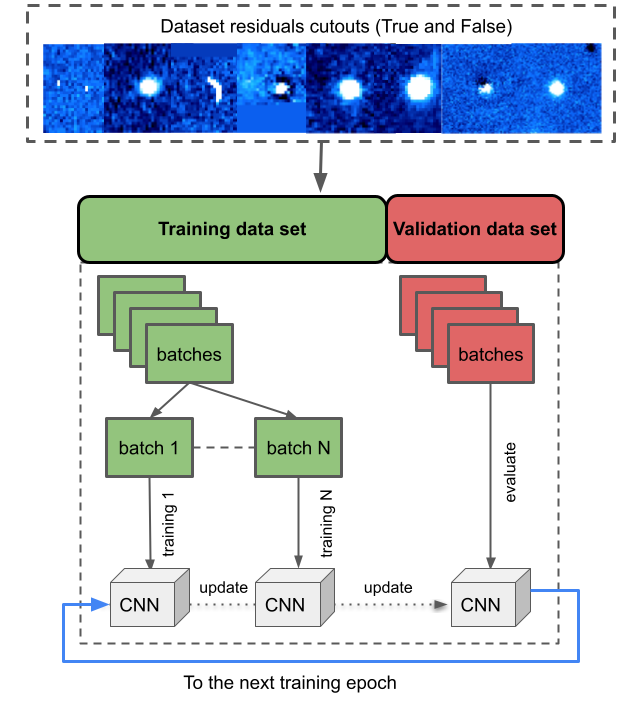}
    \caption{Schema of the training procedure in \otrain\.. In one training epoch, the original data set is split into two sub sets to train/update the CNN model and validate the classification results on a new batch of images. By repeating N times these epochs, the model converges towards the desired classification performances.}
    \label{fig:training_process}
\end{figure}

After building the CNN model, we launch the training procedure, where the model updates the parameters connecting the layers (also called trainable parameters), in order to minimize the difference between its prediction and the ground truth binary label. The data set is first ruffled and divided into training and validation sets. Each of these data sets are then split into batches, or samples. The model starts the training process by taking the first batch of images. They go through one by one and propose an update to the trainable parameters. We average these proposals and update the parameters accordingly, and move on to the next batch. After the last batch of the training  set, the updated model is applied to the validation data set, so as to see if it can perform well on images it has never seen before. This is the generalisation aspect we look for in a deep learning model. This procedure, illustrated in Figure \ref{fig:training_process}, is done within one training step also called an epoch. The more epochs, the better the model's performance will be, provided that the performance on the validation data set does not deteriorate (cf. \S\ref{subsec:overfitting}).

This procedure calls for a fine tuning of key parameters to optimize the training environment of the model. We detail them below:
\begin{itemize}
    \item \textbf{The fraction of the original data set used to make the training/validation sets: }This fraction is usually 70\%/30\% to 90\%/10\%, for the training and validation data sets respectively. It is strongly unbalanced in order to enable the model to train on the maximum number of images while keeping enough images for the validation purpose. In this work, we set this value at 85\%/15\%.
    \item \textbf{The number of epochs: } As mentioned above, this parameters influence significantly the performance of the model. We trained our model with 30 epochs, and implemented a method to trace any classification performance deterioration on the validation  data set. 
    \item \textbf{The batch size: } This parameter represents the number of images per sample in the training procedure. The bigger the size (more than 512), the more accurate our updates will be, with a trade-off with the training speed. Since we used a GPU for our training, we set this value to 1024.
    \item\textbf{The Optimizer: } The optimizer interferes in the way the trainable parameters get updated. The dominant optimizer for CNN applications is the Adam, 
    \item\textbf{The learning rate: } This configuration concerns how much the parameters get updated. We set this value to 0.001 in order to allow the model to converge to the minimum steadily.  
\end{itemize}
\subsection{Python implementation}
\begin{figure}
    \centering
    \includegraphics[width=1.0\columnwidth]{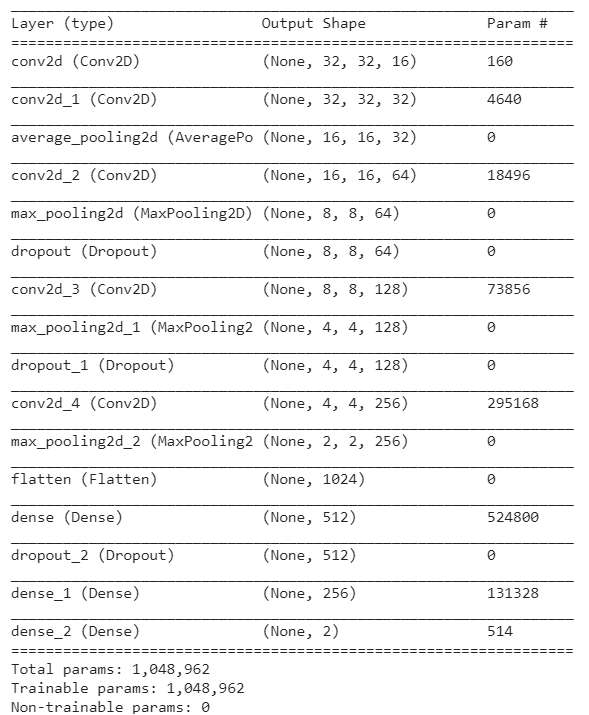}
    \caption{The layer architecture and the list of the trainable parameters of \otrain\ displayed by {\sc keras} before launching a training.}
    \label{fig:CNN_keras_arc}
\end{figure}

We implemented the model using the Python libraries {\sc tensorflow} and {\sc keras} compatible with Python version 3.7 and above. In Figure \ref{fig:CNN_keras_arc}, we show the output of {\sc keras} displaying our CNN architecture and the number of trainable parameters after launching a training procedure. All our codes as well as the diagnostic tools we present in this paper are publicly available in a documented git project \url{https://github.com/dcorre/otrain}.

\section{Training procedure of \otrain }
\label{sec:training}
\subsection{Building the training data set}
\label{sec:datasets}
To build a training data set for our \otrain\ CNN, one has to start by creating two folders \textit{True} and \textit{False} in which we store the small cutouts of the subtraction residuals centered at the positions of the real transient sources and the bogus, respectively. Each folder must contain sources that will be as close as possible to the ones the imaging instruments and the detection pipelines will output in real conditions. In addition, we require some non-standard keywords in the fits header of the \textit{True} and \textit{False} cutouts that will be used to quickly identify and characterise the candidates afterwards. We list them below:
\begin{itemize}
    \item CANDID: unique ID or name of the candidate
    \item MAG: magnitude of the source (can be optional)
    \item MAGERR: error on the magnitude of the source (can be optional)
    \item FILTER: filter used for the observation
    \item FWHM: estimation of the Full-Width at Half Maximum of the source (can be optional)
    \item EDGE: True if the source is close an image edge (depends on the detection pipeline setup). False otherwise.
\end{itemize}
The label of all the candidates stored in the \textit{True} folder is set to "1" while it is set to "0" for the candidates in the \textit{False} folder.\\
For a supervised ML algorithm as \otrain , it is required to provide a large amount of labeled candidates for the training, typically several thousands. Depending on the telescope discovery potential, OT sources are sometimes too rare in the science images to sufficiently populate the \textit{True} folder. To overcome this issue, point-like sources can be simulated in the original science images in order to artificially increase the number of real OT candidates (data augmentation techniques, see \citep{Gieseke17}). Artefacts can also be simulated if needed using softwares like \textit{SkyMaker} \citep{Bertin09}. Note that in all our analysis, we did not make use of such a technique for simulating artefacts as we had enough of them in the science and residual images after performing the image subtractions. However, we had to simulate most of our OT candidates in the original images. In the following section, we describe how we have simulated additional point-like sources in our science images and the building of  the final data cube that will then be used for training the CNN model.

\subsubsection{Point-like source simulation}
\label{sec:simulation}
To simulate point-like sources in an optical image, a model of the PSF response has to be determined to give the adequate shapes of the simulated star-like sources. We used \textit{PSFEx} to estimate the PSF response function of the science images in which we wanted to inject sources. To be more realistic, we estimated a spatially varying  PSF response for each science image in order to take into account the possible distortion of the PSF in different regions of the images and in particular close to the edges. Hence, we divided the science images into grids of $9\times 9$ regions. In each grid, we simulated N point-like sources at random positions by convolving a polynomial function with the local PSF response function. On top of these new simulated sources, we finally added a shot noise signal. Note that in our simulations, we did not take into account the positions of pre-existing sources in the science images when we inserted our simulated sources. This choice was motivated by the fact that in real conditions, an OT source could lie close to a catalogued source, be blended with it or even be detected very close to the image edges. As a consequence it may lead to possible failures when trying to detect these simulated sources depending on the detection pipeline setup.
The injected sources are simulated in a wide range of magnitudes in order to test our CNN classification performances on different conditions from bright stars up to the faintest ones close to the detection limit.

\subsubsection{The training data cube}
Once the \textit{True} and \textit{False} folders are adequately filled by enough candidate cutouts, we process all of them to build a final data cube that will be given as a single input to train our CNN model. The final data cube contains several python numpy arrays (.npy format) zipped in a dictionary-like archive as a .npz format. The different objects of the .npz dictionary store the following arrays:
\begin{enumerate}
    \item CUBE: a numpy array storing the cutouts (numpy array format)
    \item CANDIDS: a numpy array storing the unique IDs or names of the candidates 
    \item MAGS: a numpy array storing the value of the MAG keywords
    \item MAGERRS: a numpy array storing the value of the MAGERR keywords
    \item FILTERS: a numpy array storing the value of the FILTER keywords
    \item FWHMS: a numpy array storing the value of the FWHM keywords
    \item LABELS: a numpy array storing the value of the labels ("1" or "0")
\end{enumerate}
Note that if some candidates are flagged as "edge" sources, they are automatically discarded from the final data cube. Note also that the content of this datacube is not set in stone as additional relevant parameters could be added if needed. In the datacubes we have simulated, the ratio \textit{True}/\textit{False} is kept balanced (50\%/50\%) but we have implemented the possibility to produced unbalanced \textit{True}/\textit{False} data cube if needed.

\section{Testing \otrain\ on different data sets}
\label{sec:practical_cases}
One of our main goal is to propose a CNN model that can robustly classify real sources and bogus coming from a wide range of optical instruments (i.e. covering a wide range of pixel scale values, PSF response functions, depth of the limiting magnitudes, etc.) and detection pipelines. Therefore, we decided to test our \otrain\ CNN on four different telescope images processed by two different detection pipelines. Many of the science images we have used were kindly provided by the telescope teams of the GRANDMA Collaboration\footnote{\url{https://grandma.lal.in2p3.fr}} \citep{Agayeva20} which operate a world wide robotic telescope network. In particular, during the third acquisition run of the GW LIGO/Virgo detectors, GRANDMA took a large amount of images covering different sky regions \citep{Antier20a,Antier20b}. The diversity of the weather and seeing conditions found in those images allowed us to build unbiased training data sets. We choose to use the images produced by the following telescopes:
\begin{enumerate}
    \item The Javalambre Auxiliary Survey Telescope (JAST/T80) located at the Observatorio Astrofísico de Javalambre\footnote{\url{https://www.cefca.es/observatory/description}} (OAJ)
    \item The FRAM-CTA-N telescope located at the Observatorio del Roque de los Muchachos\footnote{\url{https://www.iac.es/es/observatorios-de-canarias/observatorio-del-roque-de-los-muchachos}}
    \item The TAROT Calern (TCA) telescope located at the Calern French Plateau (Observatoire de la Côte d'Azur, OCA)
    \item The TACA telescope located at Guitalens, France
\end{enumerate}
Those telescopes produce images with different pixel scales from good spatial resolution with the JAST telescope (0\farcs58/pix) up to poorly sampled images like for the TCA telescope (3\farcs31/pix). Hence, our CNN has to be robust and flexible enough to keep decent classification performances along these image feature differences.
In Table \ref{tab:telescopes}, we summarize some important properties of the images produced by these four telescopes.
\begin{table}[h!]
\caption{Telescope and image properties of the four telescopes used to test our CNN model}             
\label{tab:telescopes}      
\centering                          
\begin{tabular}{l c c c c}        
\hline\hline                 
Tel. name & D & FoV radius & pixel scale & im. size \\    
 & (cm) & (arcmin) & ("/pixel) & (pixels) \\   
\hline                        
   JAST & 83 & 60.42 & 0.58 & 9216$\times$9232 \\      
   FRAM-N & 25 & 18.72    & 1.52 & 1056$\times$1024 \\
   TACA & 16 & 51.66 & 2.81 & 1832$\times$1224 \\
   TCA & 25 & 79.81 & 3.31 & 2048$\times$2048 \\
\hline                                   
\end{tabular}
\end{table}\\
In the following sections, we briefly describe the transient detection pipelines we used to produce the inputs for \otrain\ and then, we detail the training data set we built for each telescope. 
\subsection{The {\sc gmadet} pipeline}
\label{subsec:gmadet}
The {\sc gmadet} pipeline is publicly available in the following git repository: \url{https://github.com/dcorre/gmadet}.\\
It takes science-ready images (dark, flat, bias calibrated) as inputs. First, several preprocessing steps are performed: cosmic rays are removed from this science-ready image using either {\sc lacosmic} \citep{lacosmic} or {\sc astroscrappy} \citep{astroscrappy} python package, the background is estimated using {\sc photutils} \citep{photutils} routines, the Point Spread Function is estimated using {\sc PSFEx} \citep{psfex}, finally the astrometric calibration is computed with {\sc SCAMP} using the GAIA catalog \citep{scamp}. The image is then subtracted using the {\sc HOTPANTS} code \citep{hotpants}. Following the procedure described in section \ref{sec:simulation}, we simulated N point-like sources for which we stored in an ascii file their positions in the different images. We then ran the {\sc gmadet} transient detection pipeline on those images. By cross matching the positions of the transient candidates output by {\sc gmadet} with those of the initial list of simulated point-like sources, we retrieved a vast majority of them. For these \textit{True} candidates, we produced cutouts of 32$\times$32 pixel sizes. The rest of the candidates with a failed cross match in position are directly put in the \textit{False} folder with the same cutout size (32$\times$32 pixels). With {\sc gmadet}, we usually got much more \textit{False} candidates with respect to the \textit{True} ones. To keep our final training datacube balanced, we randomly picked-up the same number of \textit{False} cutouts than in the \textit{True} folder.

\subsection{The {\sc STDPipe} pipeline}
Science-ready images are passed through a custom {\sc STDPipe} pipeline. The steps implemented in the pipeline includes extraction of objects in the image using {\sc SExtractor}, cross-matching them with the PanSTARRS DR1 catalogue, determining the photometric solution for each frame using PanSTARRS $r$ magnitude and the $g-r$ color for color correction, and image subtraction using PanSTARRS $r$ band images acquired through the {\sc HiPS2FITS} service \citep{hips2fits} as a template with the {\sc HOTPANTS} code \citep{hotpants}. Then the difference images are weighted with the image noise model, and transient detection is performed on these weighted images, taking into account the masks from both original images and templates to identify possible difference image artefacts. We simulated a set of artificial sources and injected them into the images before the image subtraction using the position-varying PSF model obtained with the PSFEx code \citep{psfex}. For the spatially coincident (within 1") transient candidate output by {\sc STDPipe} with our simulated stars, we draw some cutouts ($63\times 63$ pixels) centered at the transient candidate position and stored them in a \textit{True} folder. The rest of the transients non spatially coincident with the simulated sources are then pushed into a \textit{False} folder. They mostly consist of either a variety of CCD defects or cases where the image alignment was not good enough to properly eliminate the transients (e.g. due to not precisely exact {\sc HOTPANTS} kernel). The {\sc STDPipe} pipeline is publicly available in the git repository \url{https://github.com/karpov-sv/stdpipe}.

\subsection{The training data sets}
Below, we describe the  original images and the procedure used to build the datacubes from the four selected telescopes. The final datacube configurations for each training data set are summarized in Table \ref{tab:datacubes}.  
\subsubsection{the JAST/T80 telescope}
\begin{figure}[h!]
    \centering
    \includegraphics[width=1.0\columnwidth]{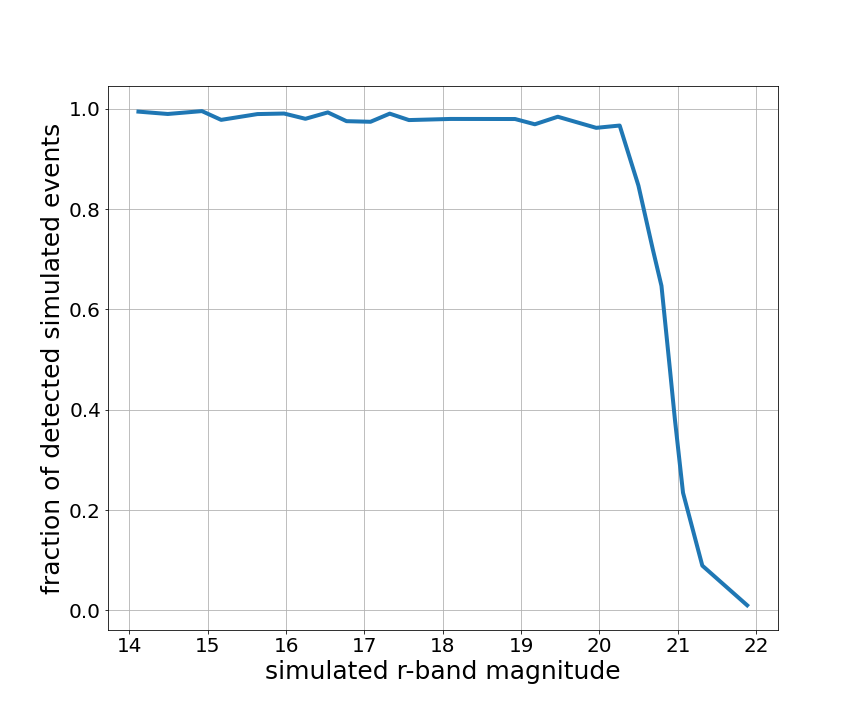}
    \caption{Fraction of the simulated point-like sources retrieved by the {\sc gmadet} transient detection pipeline as function of their simulated magnitude. The magnitudes are computed from an arbitrarily chosen zero point magnitude value. The drop observed at magnitude $\sim$ 20.5 shows the point at which the sources are close to the detection threshold (5$\sigma$).}
    \label{fig:mag_sim}
\end{figure}

We used images taken during the follow-up observations of the O3 GW event S200213t on February 2020 \citep{Blazek20,Antier20b}. After injecting artifical point-like sources in the images using both the {\sc gmadet} and the {\sc STDPipe} transient detection pipelines, we performed searches for transient candidates with the two pipelines in order to populate the \textit{True} and \textit{False} folders. Our simulated sources span a wide range of magnitudes that are drawn from an arbitrary zero point magnitude in order to cover both faint and bright transient source cases. As an example, in Figure \ref{fig:mag_sim}, we show the magnitude distribution of the simulate sources retrieved by the {\sc gmadet} pipeline.

\begin{figure}[h!]
    \centering
    \begin{minipage}{0.49\linewidth}
    \centering
    \textit{False}\par\medskip
    \includegraphics[width=0.9\textwidth]{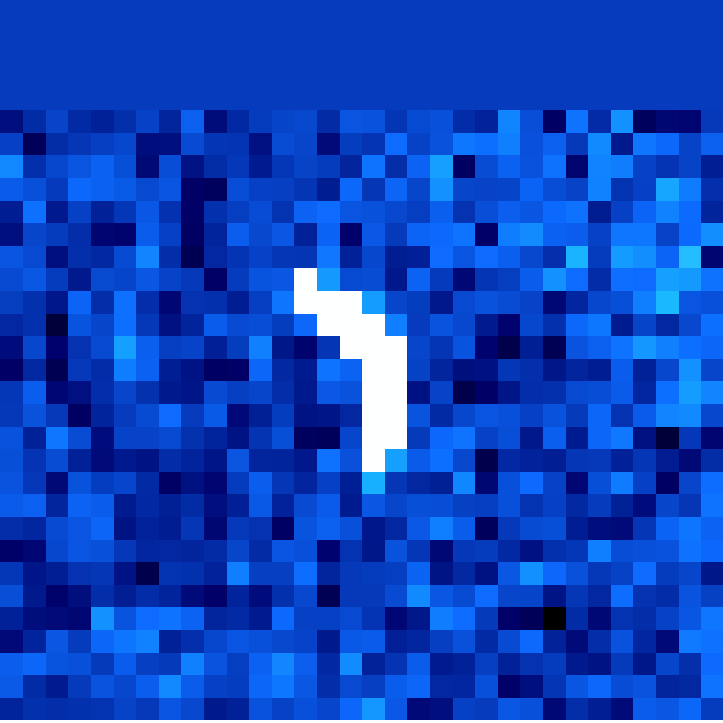}
    \end{minipage}
    \begin{minipage}{0.49\linewidth}
    \centering
    \textit{True}\par\medskip
    \includegraphics[width=0.9\textwidth]{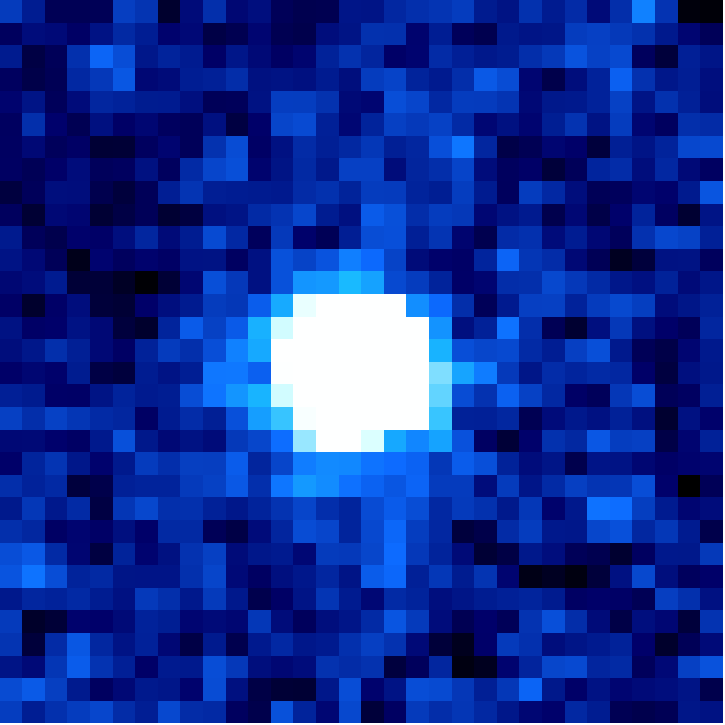}
    \end{minipage}
    \begin{minipage}{0.49\linewidth}
    \centering
    \textit{}\par\medskip
    \includegraphics[width=0.9\textwidth]{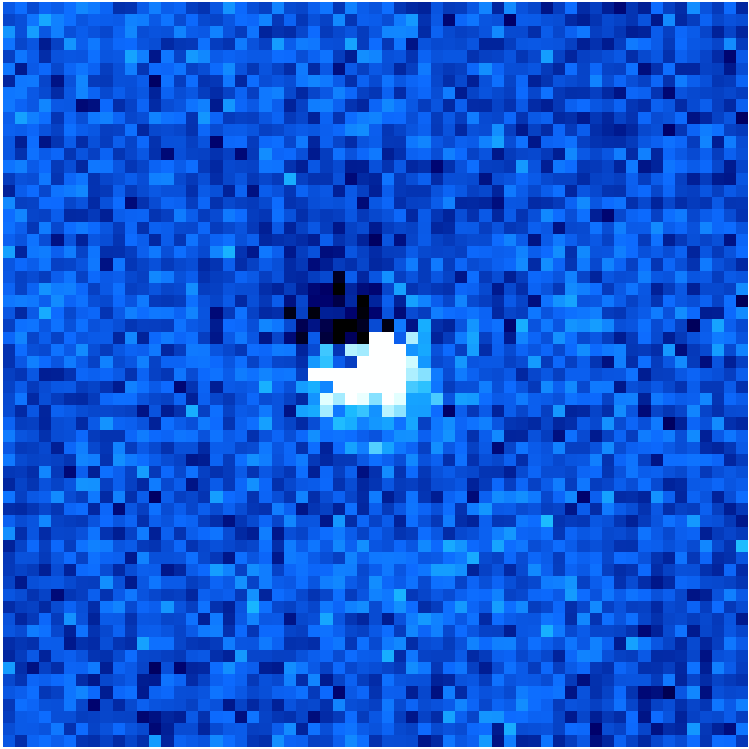}
    \end{minipage}
    \begin{minipage}{0.49\linewidth}
    \centering
    \textit{}\par\medskip
    \includegraphics[width=0.9\textwidth]{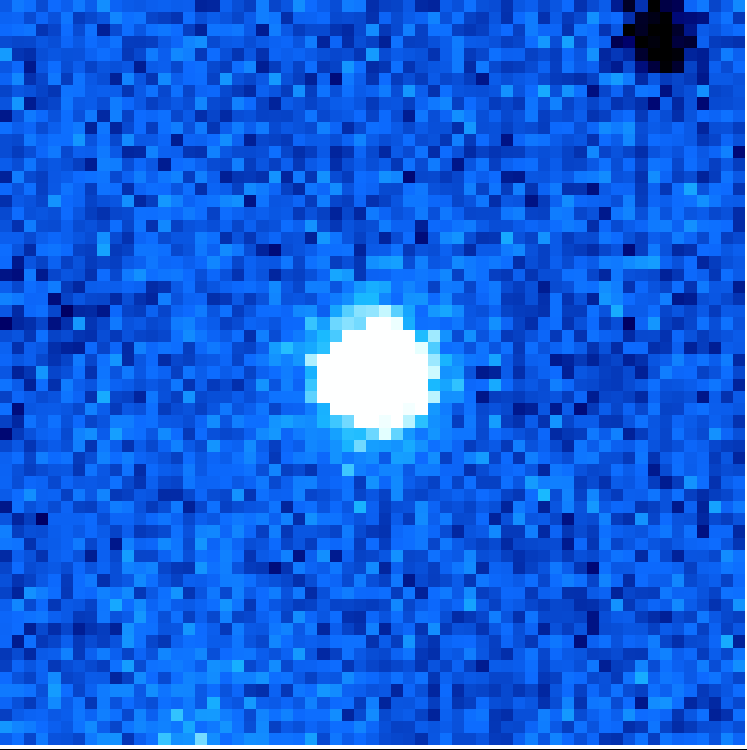}
    \end{minipage}
    \caption{Some JAST cutouts of bogus (left column) and point-like sources candidates (right column). The cutouts at the top are produced by the {\sc gmadet} pipeline (32$\times$32 pixels) while the ones at the bottom are produced by the {\sc STDPipe} pipeline (63$\times$63 pixels).}
    \label{fig:JAST_cutout}
\end{figure}

In Figure \ref{fig:JAST_cutout}, we show some examples of the residual cutouts produced by both the {\sc gmadet} and the {\sc STDPipe} pipelines and then stored in the \textit{True} and \textit{False} folders.

\subsubsection{The FRAM-CTA-N telescope}
\begin{figure}[h!]
    \centering
    \begin{minipage}{0.49\linewidth}
    \centering
    \textit{False}\par\medskip
    \includegraphics[width=0.9\textwidth]{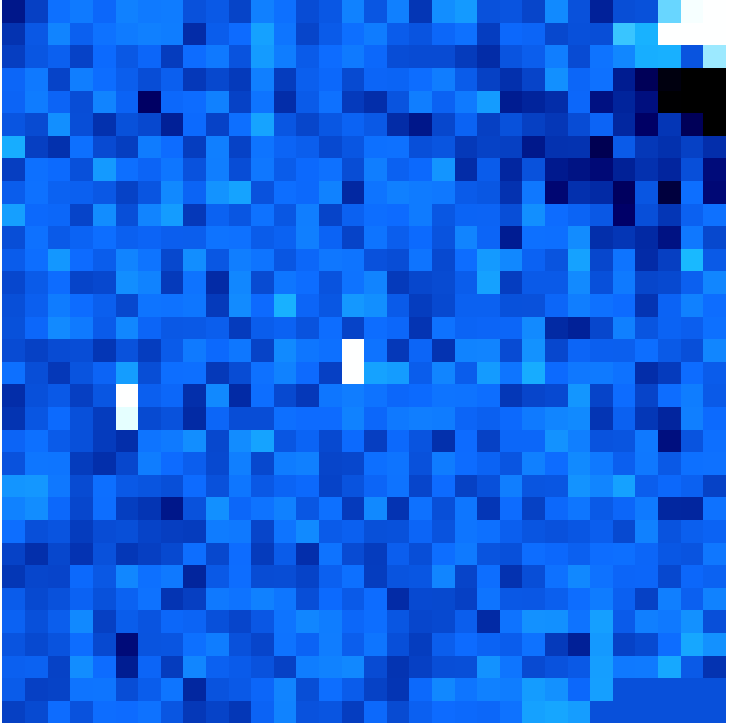}
    \end{minipage}
    \begin{minipage}{0.49\linewidth}
    \centering
    \textit{True}\par\medskip
    \includegraphics[width=0.9\textwidth]{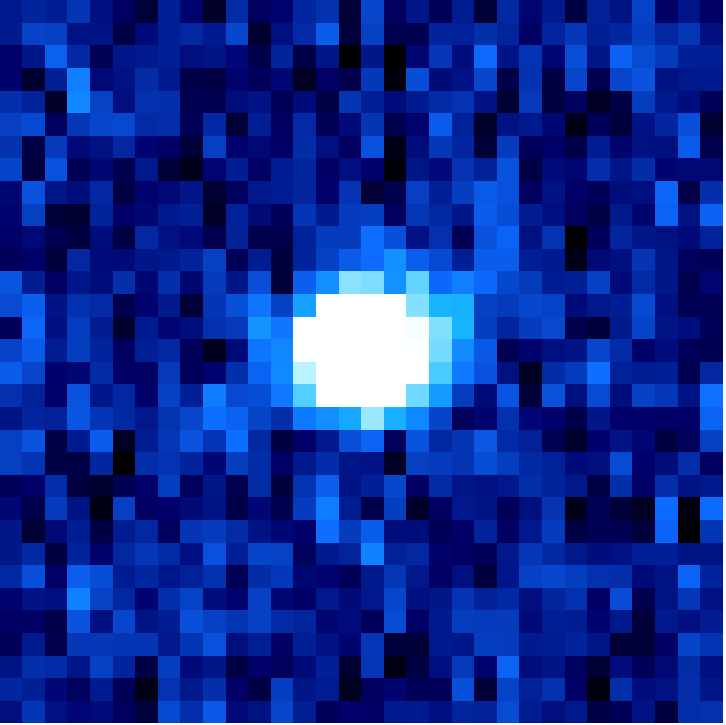}
    \end{minipage}
    \caption{Some FRAM-CTA-N cutouts of bogus (left) and point-like sources candidates (right). The cutouts at the top are produced by the {\sc gmadet} pipeline (32$\times$32 pixels).}
    \label{fig:FRAM_cutout}
\end{figure}

As a part of its nighly operation routine, the FRAM-CTA-N (a 25 cm f/6.3 telescope located at Observatorio del Roque de los Muchachos, La Palma, Spain,  equipped with $B$, $V$, $R$, and $z$ filters and having a 26$'\times$ 26$'$ field of view with 1\farcs52/pix pixel scale) telescope is performing a sky survey observations of random locations of the sky. We used the images acquired during these observations as an input for evaluating our transient classifier. To do so, the images from the telescope have been pre-processed by the telescope data acquisition and archiving system that handles initial astrometric calibration using the locally installed {\sc Astrometry.Net} code, as well as basic steps like dark subtraction, flat-fielding and masking of cosmic rays. We used the {\sc gmadet} pipeline to retrieve thousands of artificial point-like sources  we have simulated in each image. We then populated the dedicated \textit{True} and \textit{False} folders following the method described in \S \ref{subsec:gmadet}. In Figure \ref{fig:FRAM_cutout}, we show some examples of the cutouts stored in both the \textit{True} and \textit{False} folders.

\subsubsection{TACA}
\begin{figure}[h!]
    \centering
    \begin{minipage}{0.49\linewidth}
    \centering
    \textit{False}\par\medskip
    \includegraphics[width=0.9\textwidth]{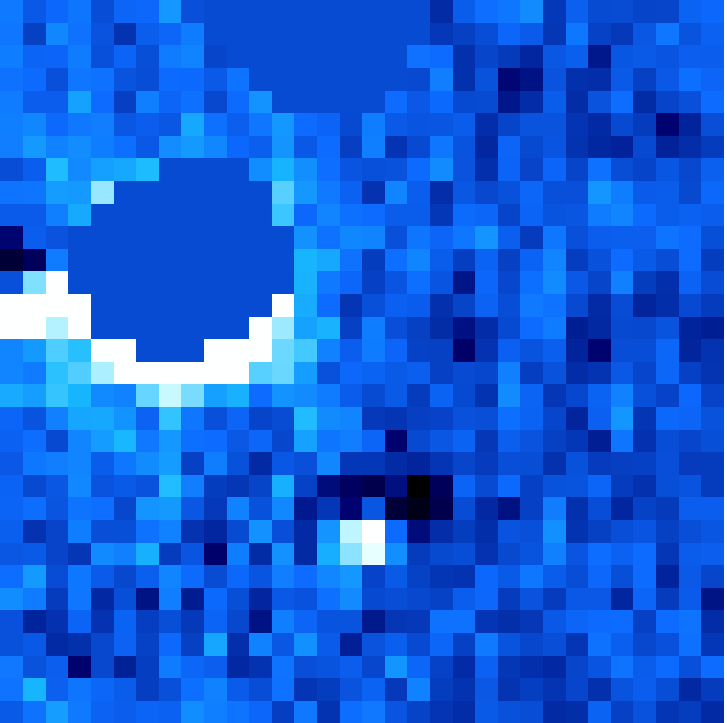}
    \end{minipage}
    \begin{minipage}{0.49\linewidth}
    \centering
    \textit{True}\par\medskip
    \includegraphics[width=0.9\textwidth]{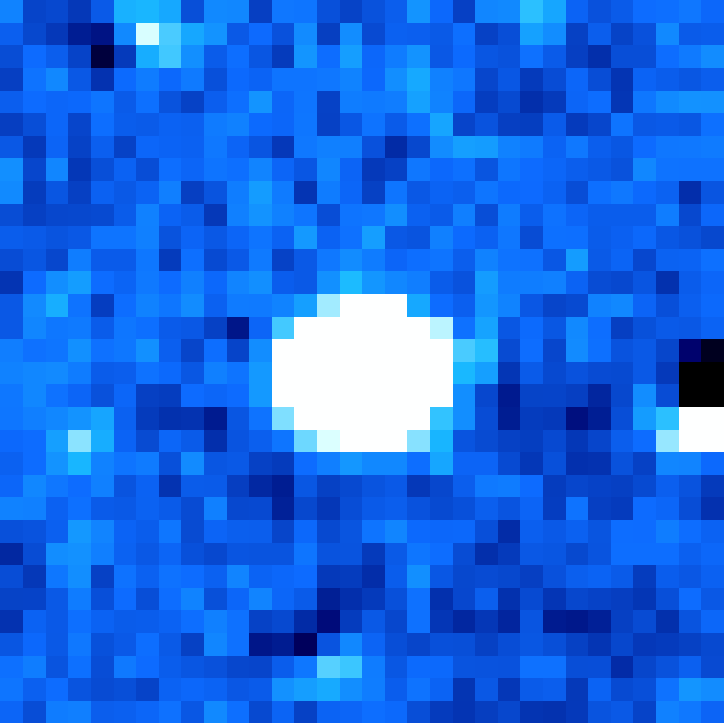}
    \end{minipage}
    \caption{Some TACA cutouts of bogus (left) and point-like sources candidates (right). The cutouts are produced by the {\sc gmadet} pipeline (32$\times$32 pixels).}
    \label{fig:TACA_cutout}
\end{figure}

We used the TACA images produced during the optical follow-up of the O3 GW event S200114f. Two nights of observations were collected on the 15-16th January 2020 to obtain 36 images of the sky. Using the {\sc gmadet} pipeline, we simulated four hundred sources per image from which we built a list of candidates. After cross-matching these candidates with the position of the simulated sources, we were able to build the \textit{True} and \textit{False} folders. In Figure \ref{fig:TACA_cutout}, we show some examples of the cutouts stored in both the \textit{True} and \textit{False} folders.

\subsubsection{TCA}
The TCA telescope took a significant number of follow-up observations during the O3 LVC campaign for the GRANDMA Collaboration \citep{Antier20a,Antier20b}. We used TCA images from various regions of the sky taken in the first half of the O3 run. We then applied the method described in \S \ref{subsec:gmadet} to simulate thousands of artificial sources in these images using the {\sc gmadet} pipeline. We finally obtained several thousand candidates to populate the datacube of \textit{True} and \textit{False} sources as shown in Table \ref{tab:datacubes}.

\begin{figure}[h!]
    \centering
    \begin{minipage}{0.49\linewidth}
    \centering
    \textit{False}\par\medskip
    \includegraphics[width=0.9\textwidth]{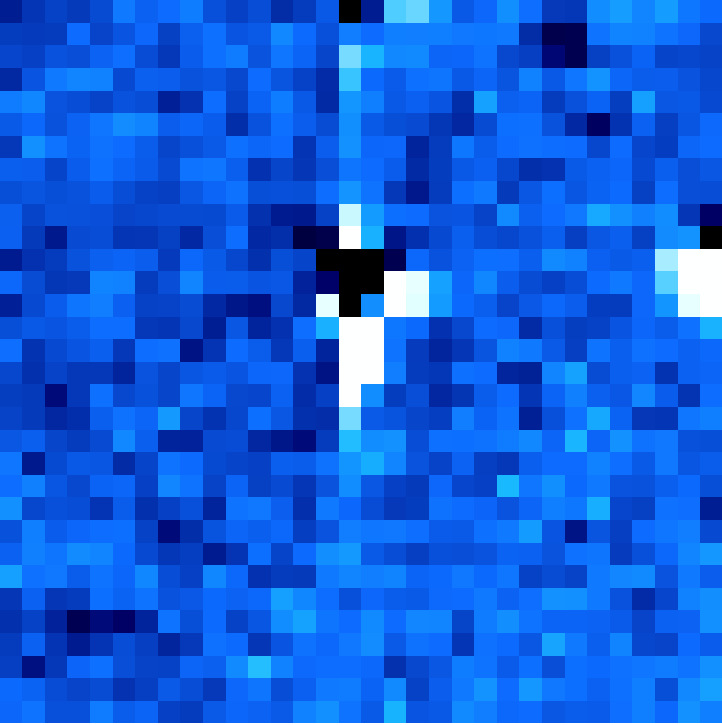}
    \end{minipage}
    \begin{minipage}{0.49\linewidth}
    \centering
    \textit{True}\par\medskip
    \includegraphics[width=0.9\textwidth]{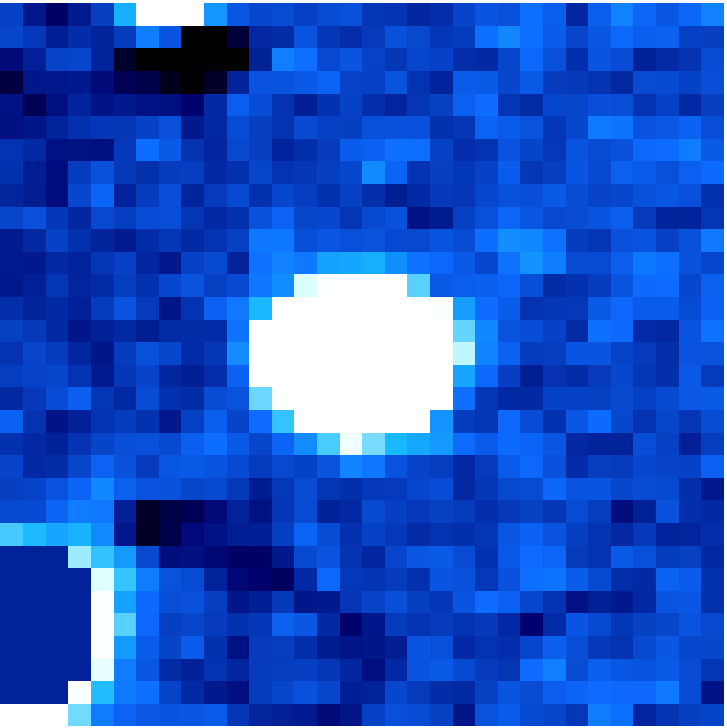}
    \end{minipage}
    \caption{Some TCA cutouts of bogus (left) and point-like sources candidates (right). The cutouts are produced by the {\sc gmadet} pipeline (32$\times$32 pixels).}
    \label{fig:TCA_cutout}
\end{figure}

In Figure \ref{fig:TCA_cutout}, we show some examples of the cutouts stored in both the \textit{True} and \textit{False} folders.

\begin{table}[h!]
\caption{Summary of the different datacube configurations used to train our \otrain\ CNN. The number of candidates reported is the sum of the labeled \textit{True} and \textit{False} candidates stored in the respective datacubes.}             
\label{tab:datacubes}      
\centering                          
\begin{tabular}{l c c c c}        
\hline\hline                 
Tel. & Detection & Number of & \textit{True}/\textit{False} & Cutouts size \\    
name & pipeline & candidates & \% & (pixels) \\   
\hline                        
   JAST & {\sc gmadet} & 13558 & 50/50 & 32$\times$32 \\      
   JAST & {\sc STDPipe} & 18328 & 50/50 & 63$\times$63 \\
   FRAM & {\sc gmadet} & 8906  & 50/50 & 32$\times$32 \\
   TACA & {\sc gmadet} & 18450 & 50/50 & 32$\times$32 \\
   TCA & {\sc gmadet} & 11414 & 50/50 & 32$\times$32 \\
\hline                                   
\end{tabular}
\end{table}
\section{Classification performance diagnosis tools}
\label{sec:tools}
\subsection{The metrics}
Once a residual cutout is analyzed by our \otrain\ CNN, it will output a probability, $\mathrm{p_{class}}$, that the source at the center is a bogus or a real source. Typically, when $\mathrm{p_{class}}=0$, the CNN model has unambiguously identified a bogus and reciprocally when $\mathrm{p_{class}}=1$, the CNN model has determined the OT candidate is real. A random guess would lead to $\mathrm{p_{class}}=0.5$. Actually, a probability threshold, $\mathrm{p_t}$, needs to be defined to take the final decision, i.e. whether a candidate is tagged as a \textit{True} or \textit{False} OT source.
In order to have a global and the most realistic perspectives of our model's performance, we implemented multiple evaluation metrics and curves. Since the threshold, $\mathrm{p_t}$, affects our metrics, we plot their values on a 2-dimensional curve with a varying threshold. This enables us to see which one works best for each telescope. Firstly, for a binary classification, we may generate the \textit{confusion matrix} to display the frequency of every combination of predicted classes and actual classes as shown in Table \ref{tab:confusion matrix}. 
\begin{table}[h!]
\centering
\caption{Components of the confusion matrix} 
\label{tab:confusion matrix}
\begin{tabular}{c || c | c }       
& Predicted & Predicted \\ 
& positive (1) & Negative (0) \\
\hline \hline    
Actual positive & True Positive  & False Negative\\
(1) & (TP) & (FN) \\
  \hline
Actual negative & False Positive\tablefootmark{a}    & True Negative  \\
(0) & (FP) & (TN) \\
\hline   
\end{tabular}
\tablefoot{
\tablefoottext{a}{Also known as False Alarm}
}
\end{table}
The confusion matrix allows to quickly identify pathological classification behaviors of our model especially if the fraction of False Positives (FP) or False Negatives (FN) is high. We typically do not want to exceed 5\% of the total candidates misclassified as FP while keeping the FN as low as possible.
The other implemented metrics help to summarize the confusion matrix and emphasize different aspects of the classification performance. We describe them below:
\begin{itemize}
    \item \textbf{Accuracy (Acc): } The percentage of candidates well classified.\\
    \item \textbf{Loss: } In our model, we use a cross entropy loss function with the following formula: 
    \begin{equation}
        \mathrm{loss = -(y\times log(p)+(1-y)\times log(1-p))}
    \end{equation} where y is a binary indicator of a class and p is the probability given to said class.\\
    \item\textbf{Precision (Prec):} Calculates the number of real point-like sources well classified by the model amongst the candidates classified as real by the model. A good precision score (near 1) shows that the model is usually right in its predictions of the positive class: Real sources.\\

    \item\textbf{Recall :} calculates how many real transients were well classified in the true transient dataset, so a good recall score indicates that the model was able to detect many positive candidates.\\

    \item\textbf{F1-score:} calculates the harmonic mean of recall and precision: 
    \begin{equation}
        \mathrm{\frac{2\times precision \times recall}{precision + recall}}
    \end{equation}
The F1-score ranges from 0 to 1, the better the performance is, the higher the value of this score.\\
    \item\textbf{Matthews correlation coefficient} \citep[MCC,][]{Matthews75}\textbf{:} Takes into account the four parts of the confusion matrix, with the following formula:
        \begin{equation}
            \mathrm{\frac{TP \times TN - FP \times FN}{\sqrt{(TP + FP)(TP + FN)(TN + FP)(TN + FN)}}}
        \end{equation}
The MCC score is in essence a measure of correlation between the observed and the predicted binary classification. They would be considered positively associated if most data falls along the diagonal cells in the confusion matrix. \\Since it's a typical correlation measure, a good MCC score is usually greater than 0.7 as it shows a good agreement between prediction and observation.\\
\end{itemize}


\begin{itemize}
    \item\textbf{ROC curve:} displays the True Positive Rate (TPR) vs the False Positive Rate (FPR), an ideal model would have a vertical line on x=FPR=0 and horizontal line for y=TPR=1. This metric enables us to visualize the global performance;
    \item\textbf{Recall-Precision curve:} like the F1-score it displays the model performance on the positive class (real sources)
\end{itemize}
The evaluation of the confusion matrix displayed by the ROC and the Recall-Precision curves, although clear and easily interpretable, might not be realistic. While the Recall-Precision curve helps us to compare the model with an always-positive classifier, it fails to include the evaluation on the negative class. On the other hand, the ROC curve leverages the four values in the confusion matrix, but its analysis could be misleading for unbalanced datasets. Even if our global datasets are balanced, we lose this property when, for example, we split the candidates into ranges of magnitude and uncertainty in magnitude. \\A more sophisticated solution was proposed by \cite{Yamashita18} to assess the performance of the model over a varying threshold, based on the MCC and the F1-score. 
\begin{itemize}
    \item\textbf{F1-MCC:} displays the MCC score versus the F1-score for a varying threshold. This allows us to determine the confidence score of the model for every telescope, by looking at the threshold that optimizes both scores. 
\end{itemize}

\subsection{Optimizing the classification threshold}
\label{sec:opt_threshold}
A threshold of $\mathrm{p_t} =0.5$ is usually the default value and the most intuitive one. But it is not always the most fitting to the model. We have conducted further studies on each telescope data set aside to decide which threshold of probability finally provides the best classifications. The ROC, precision-recall and F1-MCC curves plot respectively TPR versus FPR, precision versus recall and F1 versus MCC for a varying threshold. For the ROC curve, we select a value that maximizes the g-mean score:  
\begin{equation}
    \mathrm{g_{mean} ={\sqrt{TPR \times (1 - FPR)}}}
\end{equation}
For the precision-recall curve, we try to maximize the F1-score. 
The F1-MCC curve shows the threshold, $\mathrm{p_t}$, at which we maximize both the F1-Score and the MCC by maximizing the sum of these two metrics. The value of the best threshold differs from one telescope to another and helps us to get a better perspective of the CNN model's performance as explained in further details in the next section.

\subsection{Are we Overfitting or Underfitting? }
\label{subsec:overfitting}
Giving the model just enough data to reasonably train it is crucial. With a small dataset, the model does not have enough information to conclude the distinctions between a true transient and a bogus. The ideal case would be to have a large dataset of several tens of thousands of candidates with no contamination, and to have computational resources to be able to train a model on a dataset of this size. The model should be at least complex enough to create good non-linear connection between these characteristics, but not too complex so as not to create connections that are not relevant to the classification, and that are only present in the training data set. When the model is not trained enough, we call it an underfitting, the performance is bad on both training and validation data sets. If the model on the other hand is learning too much, and getting unimportant information, we call it an overfitting, and in this case the performance in the validation data set is considerably worse than it is on the training data set.\\
We can estimate such an over(under)fitting behavior of our CNN thanks to two metrics: the loss and the accuracy. We can track these two values throughout the training process (i.e. the epochs), to see if at some point we notice the beginning of a divergence between these values on the training and the validation data sets. If so, we should configure the model so that it stops the training at that moment.
\subsection{Understanding the model's decision with the gradient class activation map (Grad-CAM)}
\begin{figure}[h!]
    \centering
    \includegraphics[angle = 90,width=1.0\columnwidth]{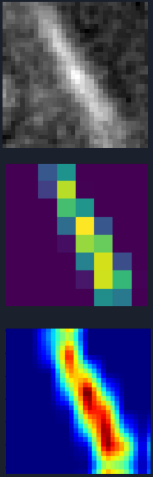}
    \caption{(Left) Residual cutout (32$\times$32 pioxels) of a bogus identified in the JAST telescope images. (Center) The GRAD-CAM Heatmap (8$\times$ 8 pixels) showing the final feature maps $\times$ their importance where our \otrain\ CNN focused on to take its decision. (Right) The rescaled Grad-CAM heat map (32$\times$32 pixels) for a direct comparison with the cutout.
 }
    \label{fig:gradcam_ex}
\end{figure}

The output of a CNN and its classification decisions are sometimes not easy to understand and to interpret. We use the Gradient-weighted Class Activation Mapping \citep[i.e. Grad-CAM,][]{Selvaraju16} to track the regions that were considered as the most important ones by the model to make its decision. The idea here is to take the last feature maps (right before flattening them) and multiply each one of them with its corresponding importance in the classification, that we get from back propagating the derivative of the loss with respect to it. We add them up and resize this output to the size of the input images in order to overlay them. The Grad-CAM is therefore a powerful diagnosis tool for the \otrain\ users to understand what triggered the CNN classification decision. In Figure \ref{fig:gradcam_ex}, we show an example of the Grad-CAM output tested on one bogus cutout from the JAST telescope. In these images, we can clearly see that the CNN model focuses on the correct region of the residual cutout to make the right decision, i.e. classifying a non point-like source as a bogus.
\section{Results and classification performances}
\label{sec:results}
Given the diagnosis tools described above, we were able to evaluate the classification performance of our CNN models trained on the five training data sets described in Table \ref{tab:datacubes}. We describe our results in the following sections.

\subsection{\otrain\ classification performance}
We applied the aforementioned metrics and curves to track the performance of the model, in both the \textit{True} and the \textit{False} classes and for a varying threshold and ranges of magnitude. This allows us to get a very general and global perspective of the performances. 
Following the method described in \S\ref{sec:opt_threshold}, we selected the best threshold, $\mathrm{p_t}$, for each telescope training set. Then we calculated the accuracy and loss of the last epoch on the validation data set, the precision, recall, F1-score and MCC as well as the confusion matrix with percentages of each part among the validation set. The results are summarized in Table \ref{tab:class_result}.

\begin{table*}[h!]
    \caption{Results of the different metrics used to evaluate the classification performances of our CNN models trained on 5 telescope datacubes. The values of the metrics have been computed based on the best probability threshold, $\mathrm{p_t}$. $\mathrm{p_t}$ has been chosen to maximize the F1-MCC scores and minimize the false alarm rate (i.e. the False Positives).}
    \centering
    \begin{tabular}{c||c|c|cccccc|c}
    \hline
        Telescope & detection &  $\mathrm{p_t}$ & Acc & Loss & Prec & Recall & F1-score & MCC & Confusion  \\
               name   & pipeline  & [0-1]      &[0-1]&[0-1] &[0-1] & [0-1]  &  [0-1]   &[0-1]&  Matrix \\
    \hline\hline
        JAST & {\sc gmadet}   & 0.35       &0.98&0.06  &0.98 & 0.99  &  0.99   &0.97 & $\begin{bmatrix}
0.49 & 0.01 \\
\le 0.01 & 0.49 
\end{bmatrix}$ \\
    \hline
        JAST & {\sc STDPipe}  & 0.51       &0.95 & 0.10  &0.94 & 0.97  &  0.96   &0.91& $\begin{bmatrix}
0.46 & 0.03 \\
0.02 & 0.5
\end{bmatrix}$ \\
    \hline
        FRAM-CTA-N & {\sc gmadet}  & 0.55  &0.93 & 0.30  &0.89 & 0.97  &  0.93   &0.85& $\begin{bmatrix}
0.43 & 0.06 \\
0.02 & 0.49
\end{bmatrix} $\\
    \hline
        TACA & {\sc gmadet}  & 0.52  &0.93 & 0.20  &0.91 & 0.96  &  0.95   &0.86& $\begin{bmatrix}
0.43 & 0.06 \\
0.01 & 0.50
\end{bmatrix}$ \\
    \hline
        TCA & {\sc gmadet}  & 0.34  &0.94 & 0.2  &0.92 & 0.97  &  0.95   &0.89& $\begin{bmatrix}
0.45  & 0.04 \\
0.01  & 0.49
\end{bmatrix}$ \\
    \hline
    \end{tabular}
    \label{tab:class_result}
\end{table*}

We pay particular attention to the MCC score since it gathers all parts of the classification (or the confusion matrix) and is more robust than other metrics. We want it to be greater than 0.7. For all of our training datacubes, the MCC scores are greater than 0.84 and therefore satisfied our scientific requirements. We also noticed that all the FP are kept below 2\% which is again perfectly in agreement with our scientific requirements (FP $\leq 5\%$).

\begin{figure*}[h!]
    \centering
    \begin{minipage}{0.49\linewidth}
    \includegraphics[trim= 0 0 0 0, clip = true,width=1.0\textwidth]{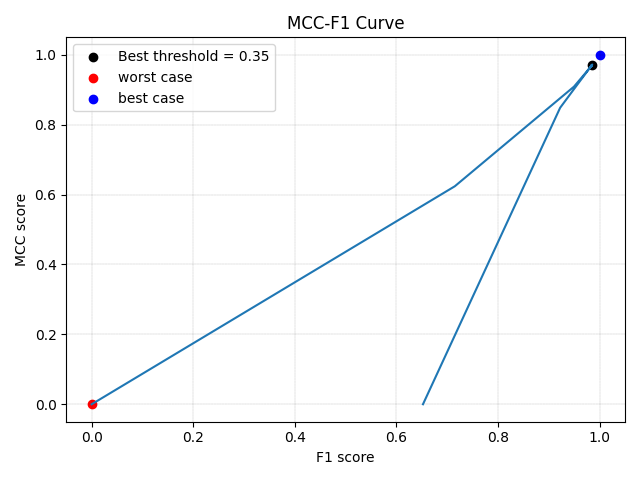}
    \end{minipage}
    \begin{minipage}{0.49\linewidth}
    \includegraphics[trim= 0 30 0 50, clip = true,width=1.0\textwidth]{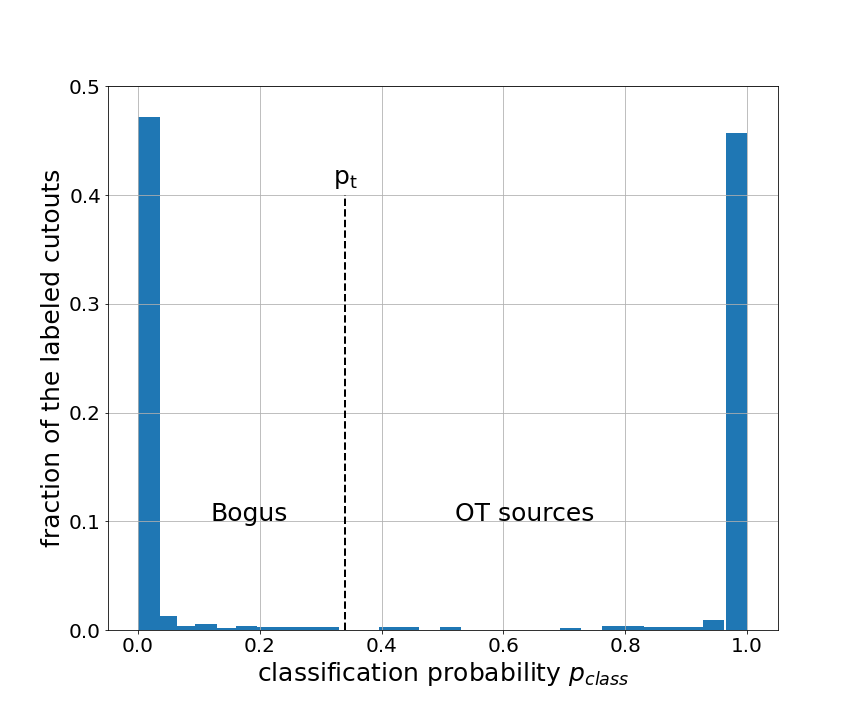}
    \end{minipage}
    \begin{minipage}{0.49\linewidth}
    \includegraphics[trim= 0 30 0 50, clip = true,width=1.0\textwidth]{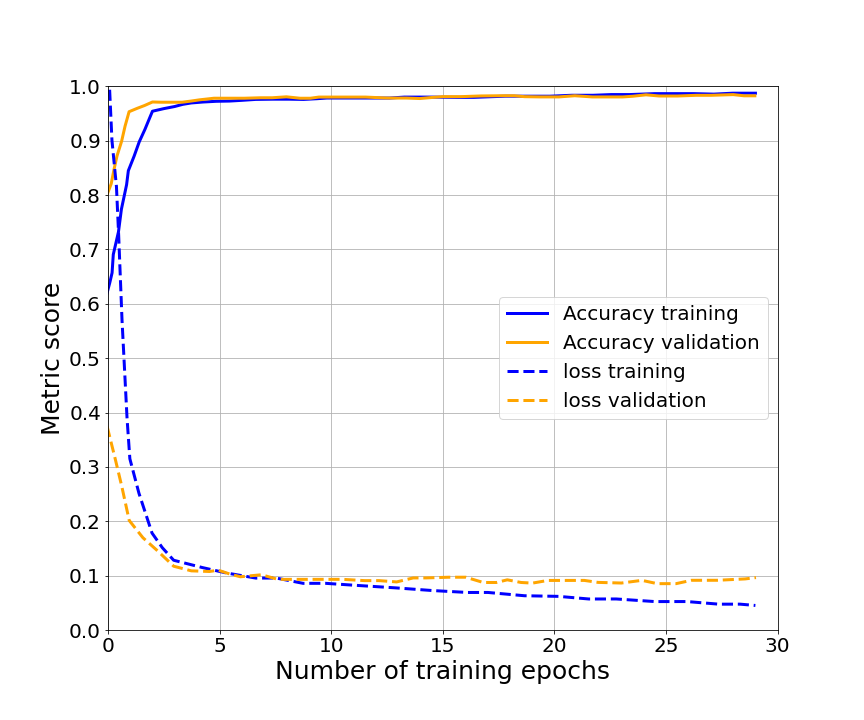}
    \end{minipage}
    \begin{minipage}{0.49\linewidth}
    \includegraphics[trim= 0 30 0 50, clip = true,width=1.0\textwidth]{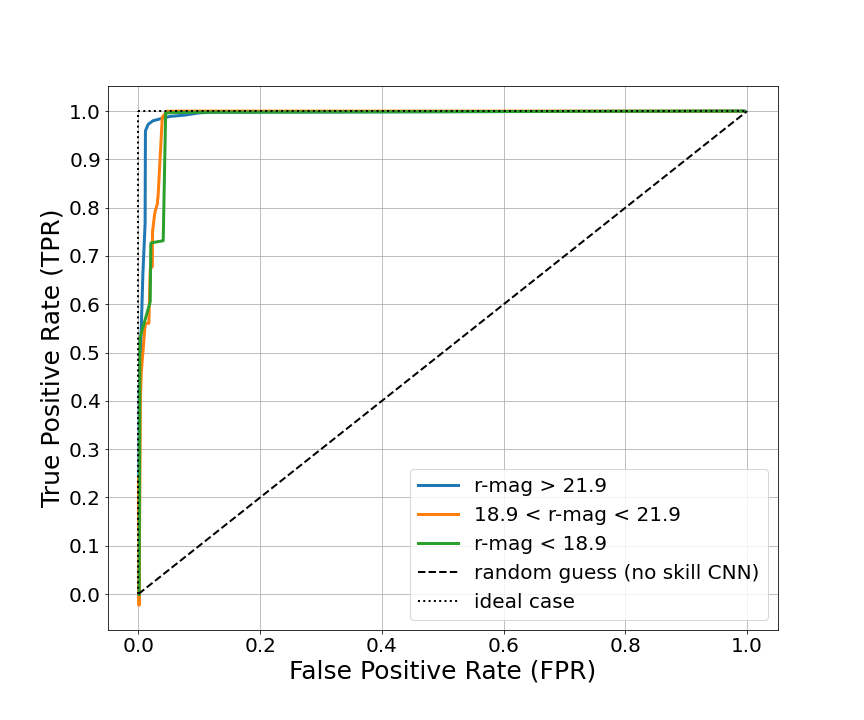}
    \end{minipage}
    \caption{Different visualization of the metrics used to evaluate the performance of the \otrain\ model trained on the JAST-{\sc gmadet} datacube. (Top left:) The evolution of the MCC and F1-score as a function of different values of $\mathrm{p_t}$. The value $\mathrm{p_t}=0.35$ maximizes both the MCC and F1-score, our two main diagnosis to evaluate the CNN classification performances. (Top right:) The classification probability distribution of the validation dataset at the last epoch of training. (Bottom left:) Evolution of the Accuracy and loss metrics as a function of the number of training epochs. (Bottom right:) The ROC curve per bin of the OT candidate magnitudes in order to diagnose the behavior of the \otrain\ model for different source brightnesses.}
    \label{fig:JAST_res}
\end{figure*}

As an example, we show in Figure \ref{fig:JAST_res}, a collection of different diagnosis curves we produced after the training of \otrain\ on the JAST cutouts created with the {\sc gmadet} pipeline.
First, we estimated the best probability threshold, $\mathrm{p_t}$ to clarify the CNN decision. Once this was fixed, we used this value ($\mathrm{p_t}=0.35$) to derive all other metrics values. On the one hand, we can see that the model unambiguously classifies the candidates, i.e. $\sim$50\% \textit{False} with $\mathrm{p_{class} = 0}$ and $\sim$50\% \textit{True} with $\mathrm{p_{class} = 1}$. Such a drastic classification behavior shows that the model is well designed for making these binary classifications and the training datacube is correctly constructed with a high purity of both \textit{True} and \textit{False} candidates. The ROC curve is near the ideal case for this telescope, and for all ranges of magnitude. These near ideal diagnosis curves associated with the metrics values listed in Table \ref{tab:class_result} allow us to validate the performance of our CNN model.  By checking at the Accuracy and the loss metrics, we note that after 10 epochs, the \otrain\ model converges towards high Accuracies (Acc$>$0.98) and small loss (loss$<0.1$). Despite a little deviation in the loss metric between the training and the validation data sets, we did not detect any significant over(under)fitting as we will show in the following section.
In the Appendix \ref{appA}, we show the same diagnosis curves for the other telescope cases we studied in this work. They all show very good classification performances in agreement with our scientific requirements.

\subsection{Evaluating the over(under)fitting behavior of the trained CNN model}
\begin{figure*}[h!]
    \centering
    \begin{minipage}{0.49\linewidth}
    \includegraphics[width=1.0\columnwidth]{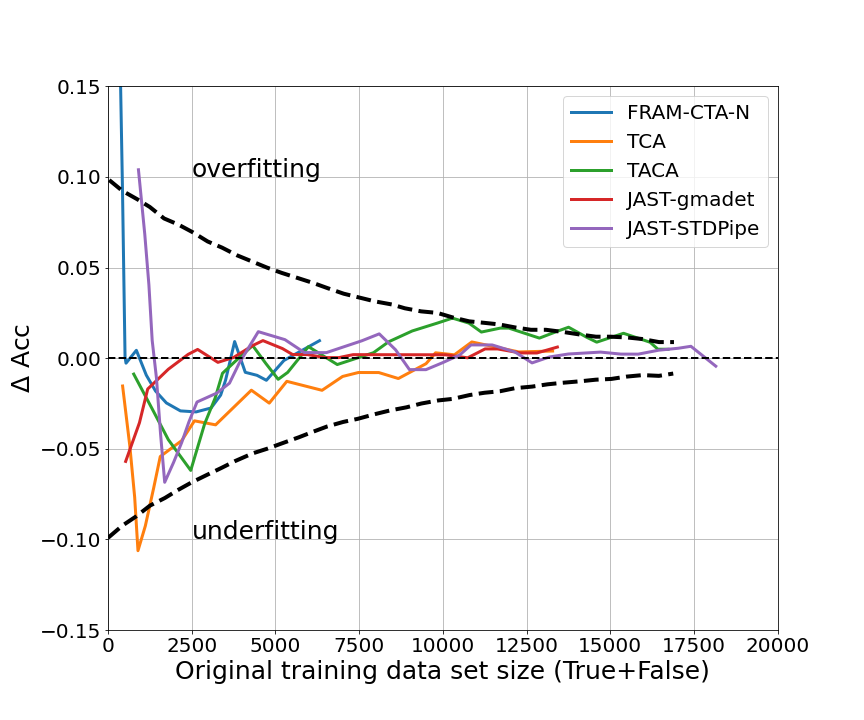}
    \end{minipage}
    \begin{minipage}{0.49\linewidth}
    \includegraphics[width=1.0\columnwidth]{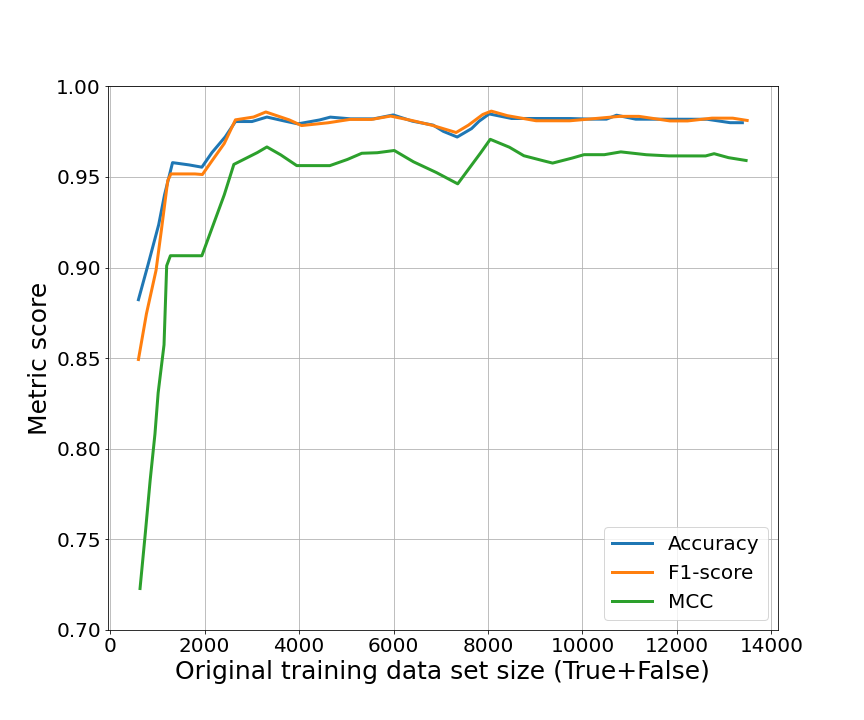}
    \end{minipage}
    \caption{ (Left:) Evolution of $\Delta Acc$ between the batch and the validation data sets as function of the size of the initial training data set after 25 epochs. This visualization allows to quickly identify if the CNN algorithm is in a over(under)fitting regime (outliers of the dashed lines). If it is, the CNN model will likely behave poorly in real conditions. (Right:) Evolution of three metrics: Accuracy, F1-score and MCC as a function of the training data set size (JAST-{\sc gmadet}).}
    \label{fig:overfitting}
\end{figure*}

As explained in section \ref{subsec:overfitting}, an important parameter that ensures the classification performances will be reproducible in real observational conditions (with a data sets never seen by our CNN model) is the over(under)fitting parameter. As we trained our CNN on relatively small data set sizes, we might be sensitive to over(under)fitting behaviors. To measure it from each of our five datacubes, we computed $\mathrm{\Delta Acc = Acc_{train} - Acc_{val}}$ as function of different data set sizes by gradually changing the sizes of the \textit{True} and \textit{False} folders used for the training. As shown in Figure \ref{fig:overfitting}, we found that with training data set sizes smaller than 5000 candidates including both \textit{True} and \textit{False} ones, the learning process leads to unavoidable over- or under-fittings with $\mathrm{|\Delta Acc|>2\%}$. For larger data set sizes, the architecture of our \otrain\ CNN is finally well adapted for being trained on such a binary classification task as it converges towards $\mathrm{\Delta Acc\sim0}$. Complementary, we also show in Figure \ref{fig:overfitting} the evolution of the Accuracy, the F1-score and the MCC metrics as functions of the data set size for the JAST-{\sc gmadet} datacube only. We also see that the metrics converge towards high scores once the dataset size used for the training is larger than 5000 cutouts.

These results allow us to conclude that our CNN model will have reproducible robust classification results while being trained on a relatively small amount of labeled cutouts (typically 10k of cutouts).

\section{Discussion and perspectives}
\label{sec:discussion}
We showed that we obtain very good RB classification results with our algorithm but there are some limitations to our work that we further discuss in this section.
\subsection{About the size of the cutouts used to train \otrain }
\begin{figure}
    \centering
    \includegraphics[width = 1.0\columnwidth]{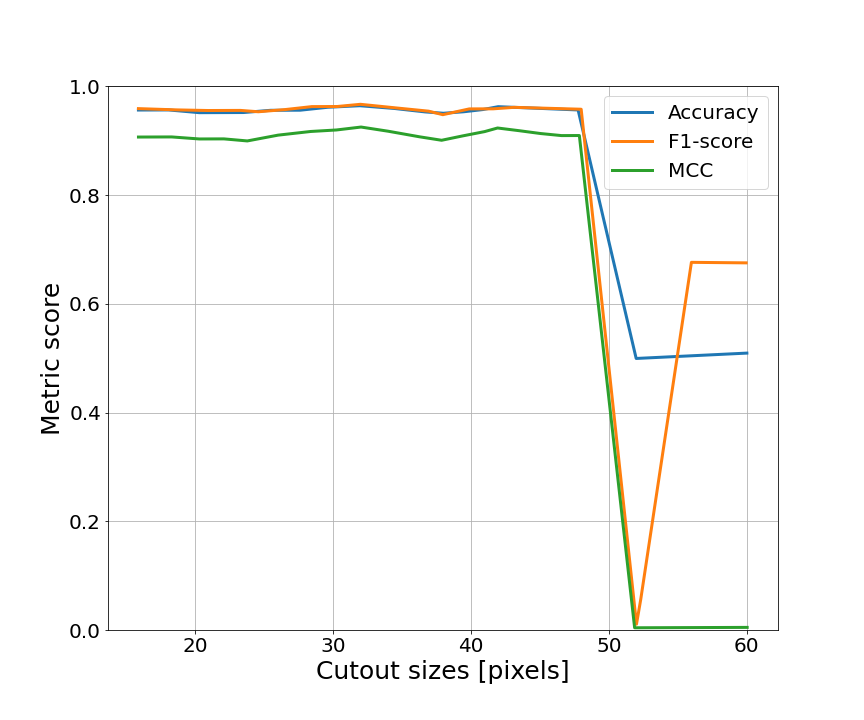}
    \caption{Evolution of the Accuracy, F1-score and MCC metric scores as a function of the cutout sizes given as input to \otrain. The architecture of our \otrain\ model applied to the JAST-{\sc STDPipe} is well adapted for a RB classification if the cutouts sizes do not exceed $\sim 50$ pixels.}
    \label{fig:cutout_size_score}
\end{figure}

The cutouts contain the informations about the OT candidate signal at the center of the image, the noise model and possibly the signal from other sources located elsewhere in the image. As the CNN model analyzes all those features in the cutouts, large cutouts may result in confusing the classification decision as more features have to be extracted and weighted. This is due to the fact that the size of the convolutional filter kernels may not be well adapated to extract numerous and large features in the cutouts. Therefore, we decided to study the impact of the cutout sizes in our CNN decisions. To do so, we used the cutouts (\textit{False} and \textit{True}) produced by the {\sc STDPipe} pipeline in 63$\times$63 pixels from the JAST telescope original images. We built several datacubes with the cutout sizes varying from $15\times15$ pixels to $63\times 63$ pixels while still keeping the OT candidates at the center. We show, in Figure \ref{fig:cutout_size_score}, that the maximum acceptable size for these cutouts to keep high classification performances are 48$\times$48 pixels. Above this cutout size, the model behaves randomly and badly classifies the candidates. We attribute this to a structural limitation of our CNN algorithm which employs small kernel filters (3$\times$3 pixels) in the convolutional layers. These kernels might therefore not be adapted to catch features in cutouts larger than 48 pixels.
\subsection{Can we use more information to improve the classification ?}
Currently, our CNN algorithm performs its classification process on the residual cutouts provided by the subtraction of science and reference images. While we obtained few False Positives, we ended with a non negligible number of False Negatives. In other words, we optimized \otrain\ to keep the false alarm rate as low as possible with the cost of losing real optical transient sources (conservative approach). However, additional informations on each candidate could be used to reduce the FN and hence improve the classification performances. \cite{Gieseke17} explored this approach by considering three cases for their CNN training:
\begin{enumerate}
    \item the CNN predictions are based on the analysis of the science, the reference and the residual cutouts
    \item the CNN predictions are based on the analysis of the science and the reference cutouts
    \item the CNN predictions are based on the analysis of the residual cutouts only.
\end{enumerate}
According to their study, they concluded that the training data sets based on the first and second configurations described above give similarly good classification performances. On the contrary, their CNN model gave slightly worse results when it only analyzed the residual cutouts. An interesting perspective is highlighted here, i.e. the possibility of discarding the image subtraction process (that can produce plenty of artefacts) for RB classification purposes. In our approach, we only used the standard residual images output from the {\sc gmadet} and {\sc STDPipe} pipelines which already provides decent classification results. In order to give more flexibility and scientific perspectives to \otrain, an update allowing the users to give different types of cutouts as a datacube input (science-ready, reference, residual, mask template images) is foreseen.

\subsection{Towards more sophisticated methods : FASTER-R and MASK-R CNN}

\begin{figure}[h!]
    \centering
    \includegraphics[width=0.7\columnwidth]{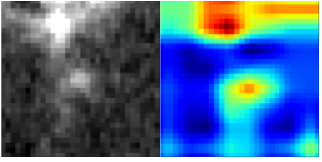}
    \caption{Example of a misclassified OT (FN) by \otrain. While the CNN focuses a fraction of its attention into the central part of the cutouts (where the OT is), most of its classification decision relies on the presence of a brighter source and a non-uniform background noise at the top of the cutouts.}
    \label{fig:FN_GRAD-CAM}
\end{figure}

Thanks to the Grad-CAM tools included in \otrain\,, we noticed that some OT candidates were misclassified due to the presence of additional sources in the cutouts, see Figure \ref{fig:FN_GRAD-CAM}.

This kind of misclassification might be solved by the addition of more informations in the datacube as discussed above. It could also be completely avoided by using novel techniques to analyze the astronomical images by deep convolutional neural networks such as: FASTER-R and/or MASK-R CNN algorithms. \\
Whilst both of them consists in finding regions in the image that contains the object (i.e OT) we are looking for, The FASTER-R CNN will output bounding boxes containing these objects whereas the MASK-R-CNN goes one step further and gives us the exact pixels of the said objects. \\

For the training, the FASTER-R-CNN as explained by \cite{DBLP:journals/corr/RenHG015}, takes on a dataset of images with information such as the coordinates of the bounding boxes containing sources of the image. It applies a CNN model to extract features, and passes them to another model called "Region Proposal Network" (RPN) which performs a binary classification (presence of an object) and a regression on the coordinates. Like all training processes, the model tries to align its predictions with the ground truth. In this case, it places a sliding window (anchors) that could vary in size and proportions and then calculates the score of  Intersection over the Union (IoU) between its predicted boxes and the ones in the dataset we had given it.\\ The RPN will output several proposals, for all objects detected in the image. Since these regions will not have the same size, a pooling will be applied before passing it through a classifier, that will have as input the feature maps of these parts of the original image (i.e the regions of interest), instead of the whole image at once. The MASK-R CNN outperforms the FASTER-R-CNN as it adds to this structure a branch for a binary mask, showing if the pixel is part of the object or not, thus performing a pixel-level detection of objects. \cite{DBLP:journals/corr/HeGDG17} noticed a misalignment between the ROIs after the pooling layer, and the regions in the original image, this could cause a drop in the model's performance, since the mask requires precision at pixel level. They adjust this part, by using ROIAlign instead of ROIPool, where they apply a bilinear interpolation on the grid points of the feature maps in order to get the exact values in pooled regions.\\ This model could be well adapted to our RB problem, since some astronomical tools, like {\sc SExtractor}, exist to create a segmentation map of the optical sources present in the image.

While implementing these new classification methods were beyond the scope of this paper, further development will be engaged to include them in the \otrain\ framework.

\section{Conclusions}
\label{sec:conclusion}
Real/bogus classifiers are now standard in the pipelines aiming at detecting a large number of transient phenomena at optical wavelengths. We have developed a robust and flexible convolutional neural network model called \otrain. It aims at distinguishing real point-like sources from various types of bogus in optical images. We have shown that \otrain\ reaches high classification performances for a wide range of telescope pixel scales and observational conditions. In addition, we have demonstrated the capabilities of our CNN model to robustly behave against various types of inputs (cutout and dataset sizes, residuals images from different subtraction methods, etc.) provided by two publicly available transient detection pipelines ({\sc gmadet} and {\sc STDPipe}). Indeed, on the five training datacubes, we obtained very good classification performances with MCC scores ranging in the interval [0.84 - 0.99]. Such performances can be obtained with a relatively small size of the training data sets, typically few tens of thousands of labeled cutouts, without creating any over(under)fitting during the training epochs. To guide the users, we have built a complete user friendly and easy-to-use frame work to launch training as well as to check diagnostic tools monitoring the performance of the CNN model. This will greatly help observers who would like to use such a machine learning technique in their own pipeline while being not fully experts. \otrain\ is a publicly available code and aims to be upgradeable in the future with new features that would enlarge the flexibility of the code and the classification perspectives such as the simultaneous classification of multiple sources at once.
   
\begin{acknowledgements}
      DT is funded by the CNES Postdoctoral Fellowship program at the CEA-Saclay/AIM/Irfu laboratory. 
      SK acknowledges support from the European Structural and Investment Fund and the Czech Ministry of Education, Youth and Sports (Project CoGraDS -- CZ.02.1.01/0.0/0.0/15\_003/0000437).
      FRAM-CTA-N operation is supported by the Czech Ministry of Education, Youth and Sports (projects LM2015046, LM2018105, LTT17006) and by European Structural and Investment Fund and the Czech Ministry of Education, Youth and Sports (projects CZ.02.1.01/0.0/0.0/16\_013/0001403 and CZ.02.1.01/0.0/0.0/18\_046/0016007).
      DAK acknowledges support from Spanish National Research Project RTI2018-098104-J-I00 (GRBPhot).
      This research has made use of the VizieR catalogue access tool, CDS, Strasbourg, France. We also want to thank Pascal Yim, professor at Ecole Centrale Lille, for his inputs on Faster-R-CNN and Mask-R-CNN models.  The authors warmly thank the GRANDMA Collaboration and Alain Klotz for providing images taken from their optical follow-up of gravitational wave sources during the O3 run.
\end{acknowledgements}

%
%

\bibliographystyle{aa}
\bibliography{biblio}

\begin{appendix}
\section{\otrain\ diagnosis curves for JAST-{\sc STDPipe}, TACA, FRAM and TCA images}
\label{appA}
\begin{figure*}[h!]
    \centering
    \textbf{JAST-{\sc STDPipe} post-training diagnosis curves}\\
    \begin{minipage}{0.49\linewidth}
    \includegraphics[trim= 0 0 0 0, clip = true,width=1.0\textwidth]{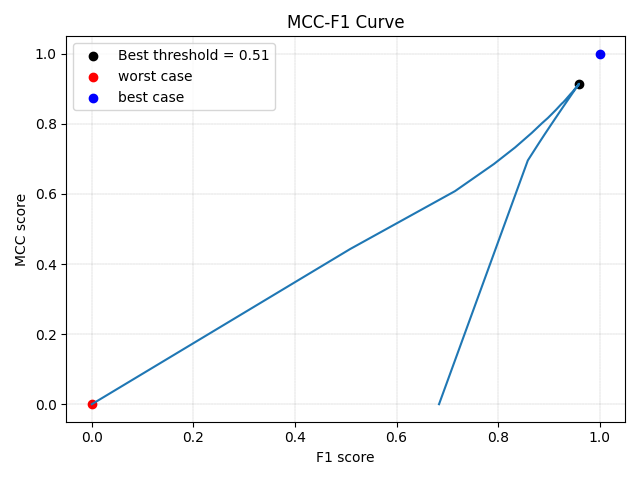}
    \end{minipage}
    \begin{minipage}{0.49\linewidth}
    \includegraphics[trim= 0 30 0 50, clip = true,width=1.0\textwidth]{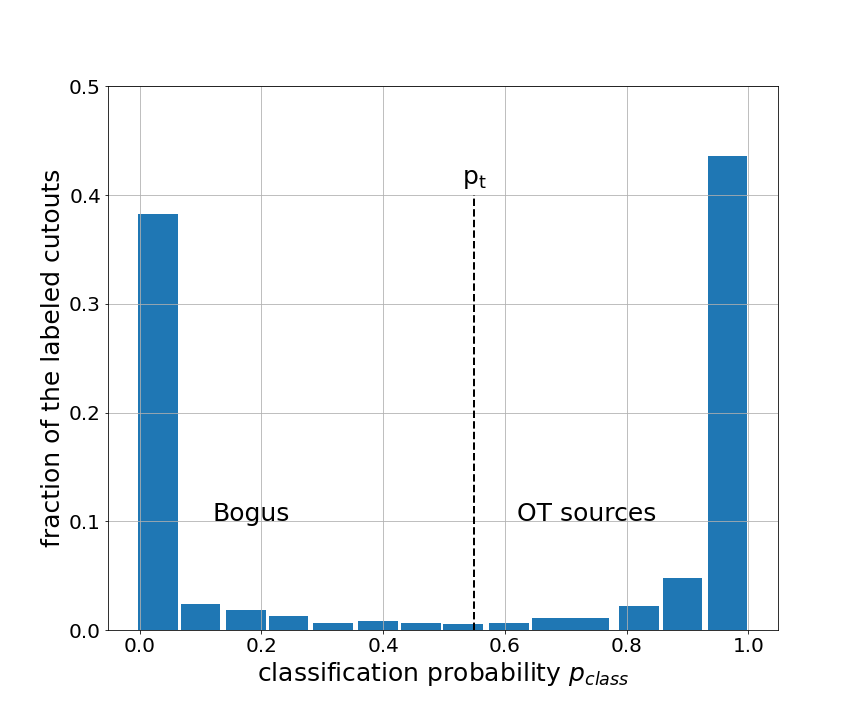}
    \end{minipage}
    \begin{minipage}{0.49\linewidth}
    \includegraphics[trim= 0 30 0 50, clip = true,width=1.0\textwidth]{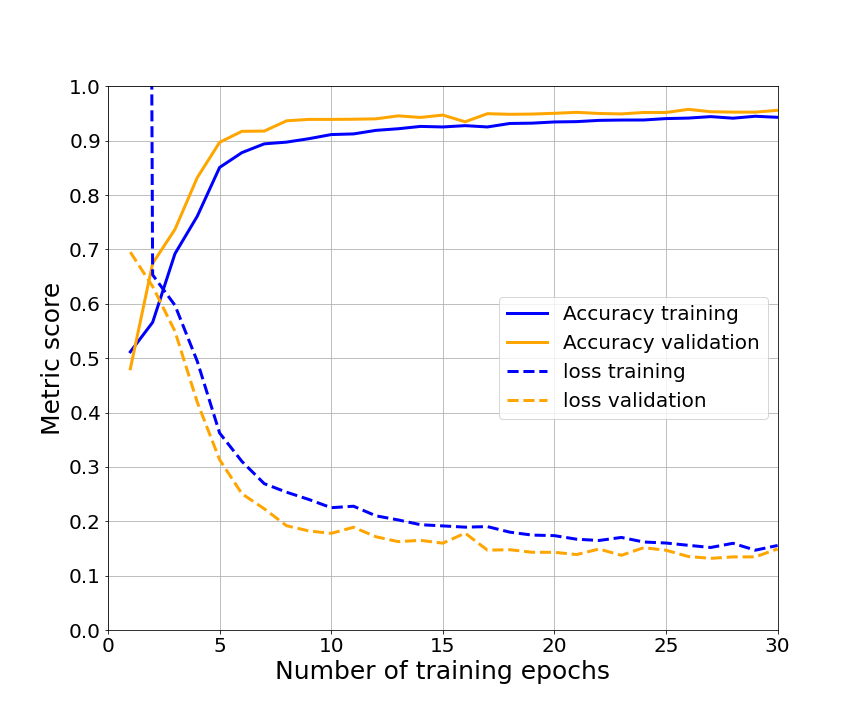}
    \end{minipage}
    \begin{minipage}{0.49\linewidth}
    \includegraphics[trim= 0 30 0 50, clip = true,width=1.0\textwidth]{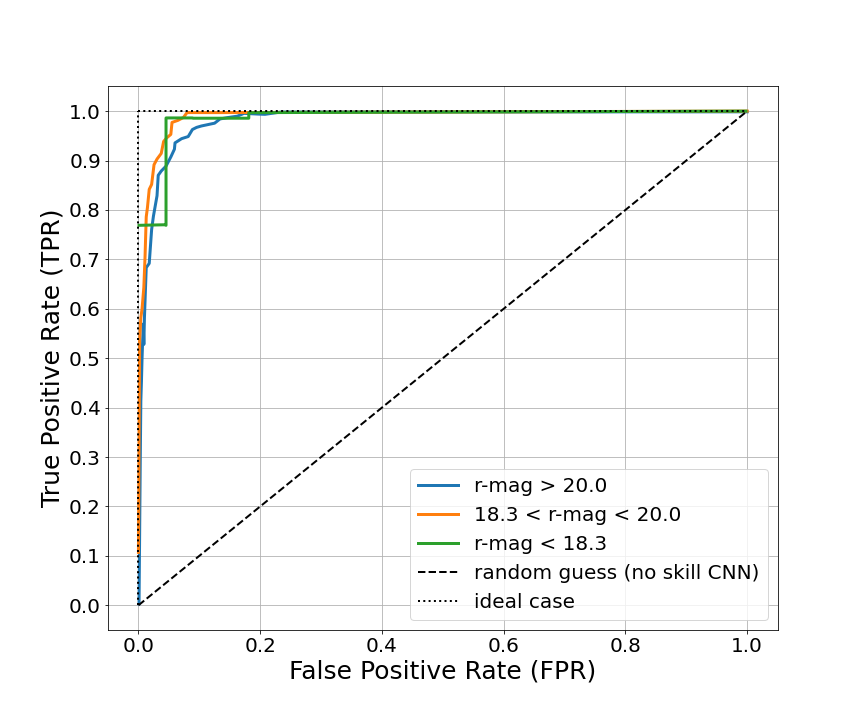}
    \end{minipage}
    \caption{Different visualization of the metrics used to evaluate the performance of the \otrain\ model trained on the JAST-{\sc STDPipe} datacube. (Top left:) The evolution of the MCC and F1-score as a function of different values of $\mathrm{p_t}$. The value $\mathrm{p_t}=0.51$ (best threshold) maximizes both the MCC and F1-score. (Top right:) The classification probability distribution of the validation dataset at the last epoch of the training. (Bottom left:) Evolution of the Accuracy and loss metrics as a function of the number of training epochs. (Bottom right:) The ROC curve per bin of the OT candidate magnitudes in order to diagnose the behavior of the \otrain\ model for different source brightnesses.}
    \label{fig:JAST-stdpipe_res}
\end{figure*}
\begin{figure*}[h!]
    \centering
    \textbf{TACA post-training diagnosis curves}\\
    \begin{minipage}{0.49\linewidth}
    \includegraphics[trim= 0 0 0 0, clip = true,width=1.0\textwidth]{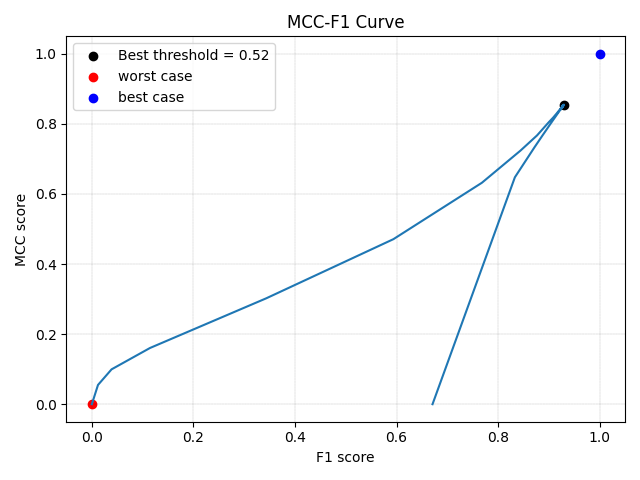}
    \end{minipage}
    \begin{minipage}{0.49\linewidth}
    \includegraphics[trim= 0 30 0 50, clip = true,width=1.0\textwidth]{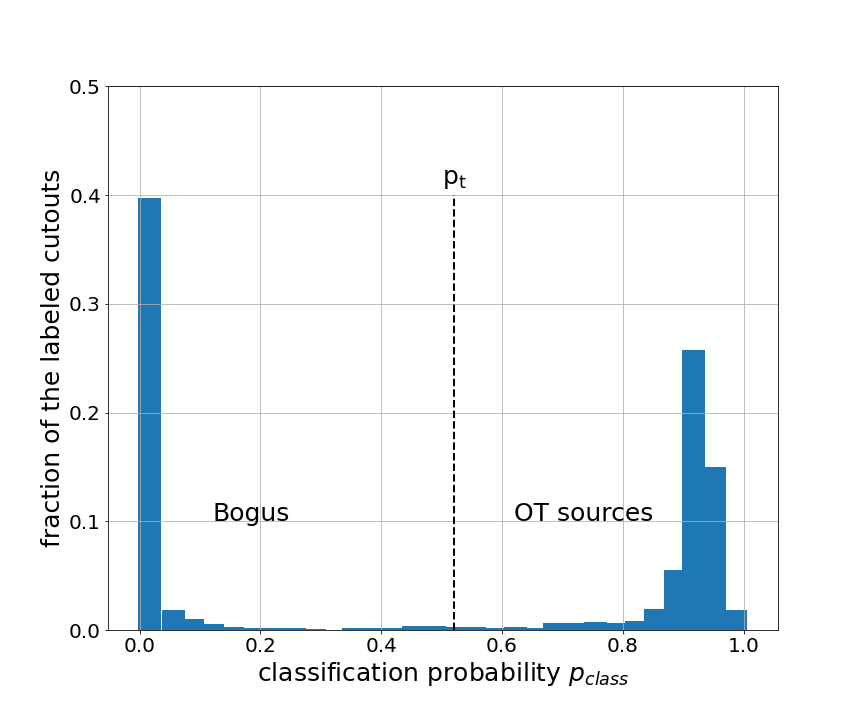}
    \end{minipage}
    \begin{minipage}{0.49\linewidth}
    \includegraphics[trim= 0 30 0 50, clip = true,width=1.0\textwidth]{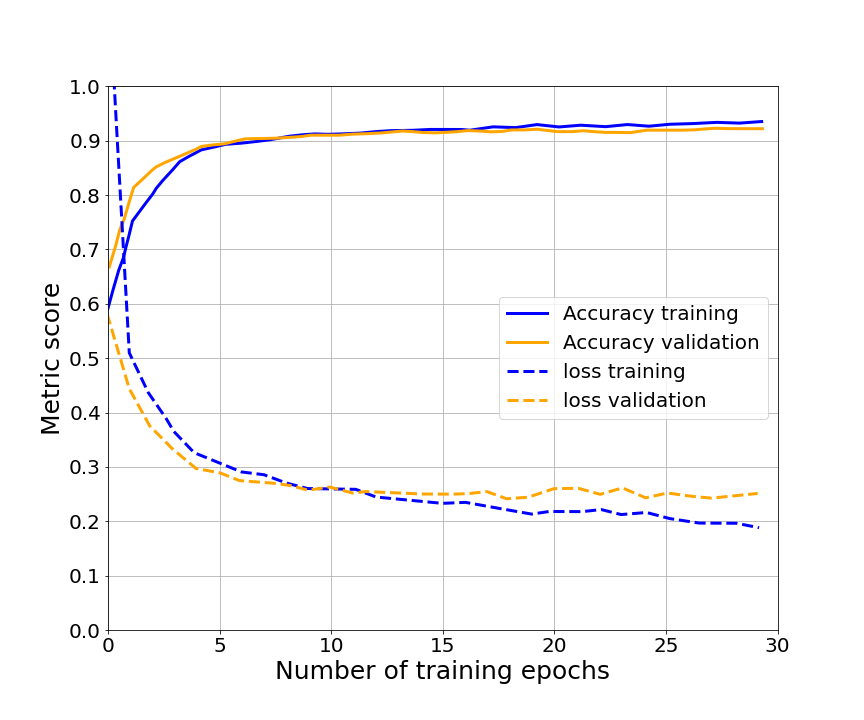}
    \end{minipage}
    \begin{minipage}{0.49\linewidth}
    \includegraphics[trim= 0 30 0 50, clip = true,width=1.0\textwidth]{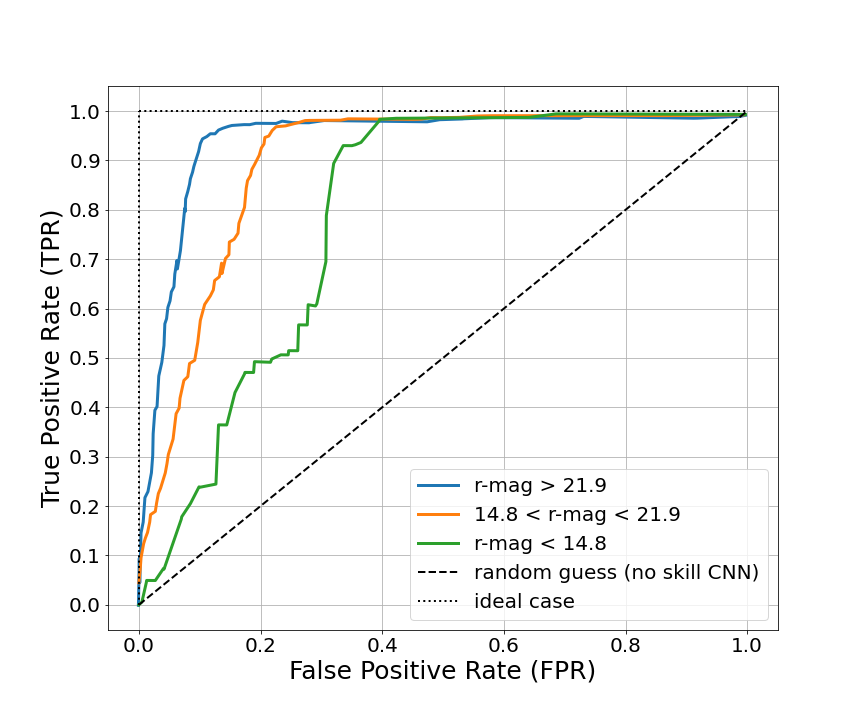}
    \end{minipage}
    \caption{Different visualization of the metrics used to evaluate the performance of the \otrain\ model trained on the TACA-{\sc gmadet} datacube. (Top left:) The evolution of the MCC and F1-score as a function of different values of $\mathrm{p_t}$. The value $\mathrm{p_t}=0.52$ (best threshold) maximizes both the MCC and F1-score. (Top right:) The classification probability distribution of the validation dataset at the last epoch of the training. (Bottom left:) Evolution of the Accuracy and loss metrics as a function of the number of training epochs. (Bottom right:) The ROC curve per bin of the OT candidate magnitudes in order to diagnose the behavior of the \otrain\ model for different source brightnesses.}
    \label{fig:TACA_res}
\end{figure*}
\begin{figure*}[h!]
    \centering
    \textbf{FRAM-CTA-N post-training diagnosis curves}\\
    \begin{minipage}{0.49\linewidth}
    \includegraphics[trim= 0 0 0 0, clip = true,width=1.0\textwidth]{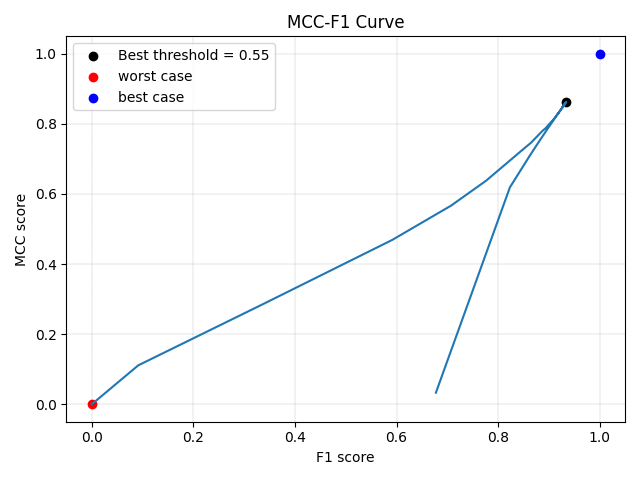}
    \end{minipage}
    \begin{minipage}{0.49\linewidth}
    \includegraphics[trim= 0 30 0 50, clip = true,width=1.0\textwidth]{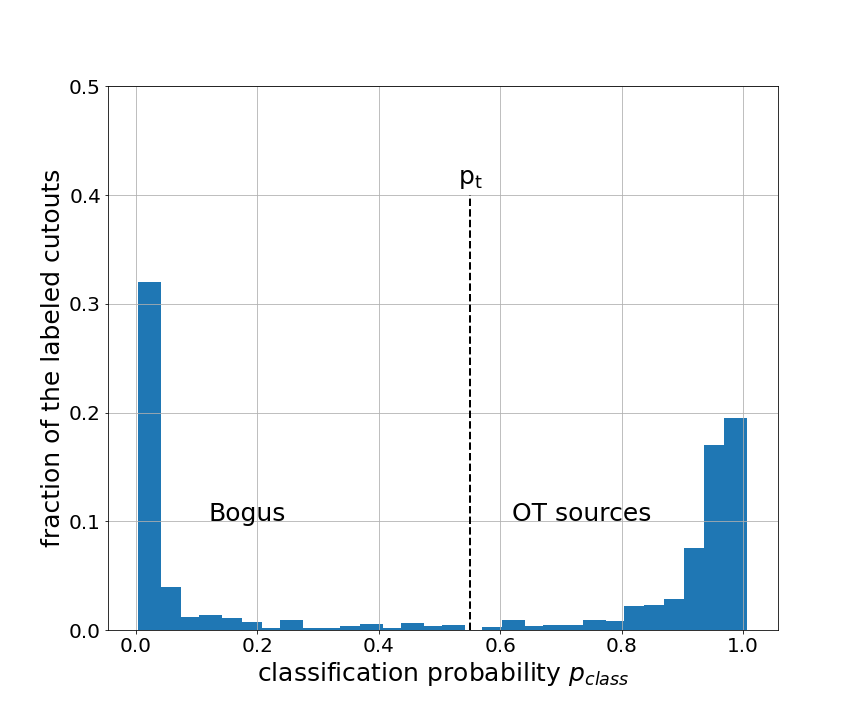}
    \end{minipage}
    \begin{minipage}{0.49\linewidth}
    \includegraphics[trim= 0 30 0 50, clip = true,width=1.0\textwidth]{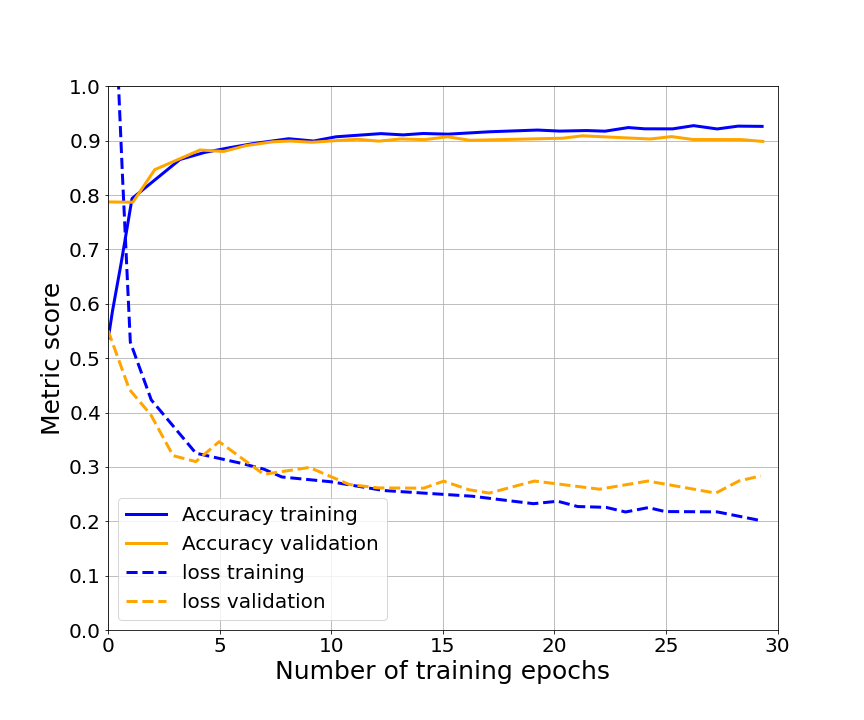}
    \end{minipage}
    \begin{minipage}{0.49\linewidth}
    \includegraphics[trim= 0 30 0 50, clip = true,width=1.0\textwidth]{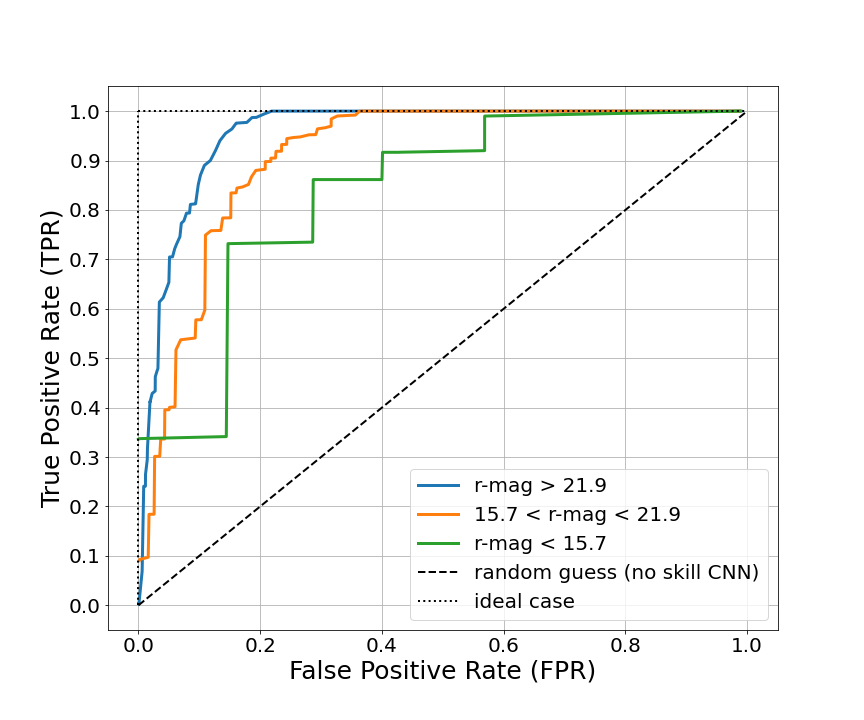}
    \end{minipage}
    \caption{Different visualization of the metrics used to evaluate the performance of the \otrain\ model trained on the FRAM-CTA-N-{\sc gmadet} datacube. (Top left:) The evolution of the MCC and F1-score as a function of different values of $\mathrm{p_t}$. The value $\mathrm{p_t}=0.55$ (best threshold) maximizes both the MCC and F1-score. (Top right:) The classification probability distribution of the validation dataset at the last epoch of the training. (Bottom left:) Evolution of the Accuracy and loss metrics as a function of the number of training epochs. (Bottom right:) The ROC curve per bin of the OT candidate magnitudes in order to diagnose the behavior of the \otrain\ model for different source brightnesses.}
    \label{fig:FRAM_res}
\end{figure*}
\begin{figure*}[h!]
    \centering
    \textbf{TCA post-training diagnosis curves}\\
    \begin{minipage}{0.49\linewidth}
    \includegraphics[trim= 0 0 0 0, clip = true,width=1.0\textwidth]{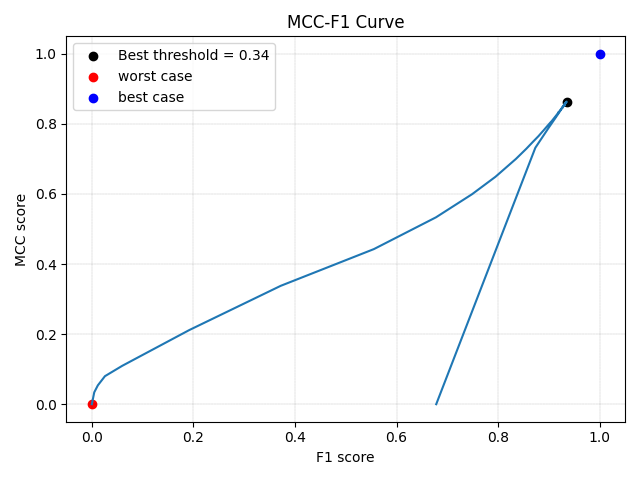}
    \end{minipage}
    \begin{minipage}{0.49\linewidth}
    \includegraphics[trim= 0 30 0 50, clip = true,width=1.0\textwidth]{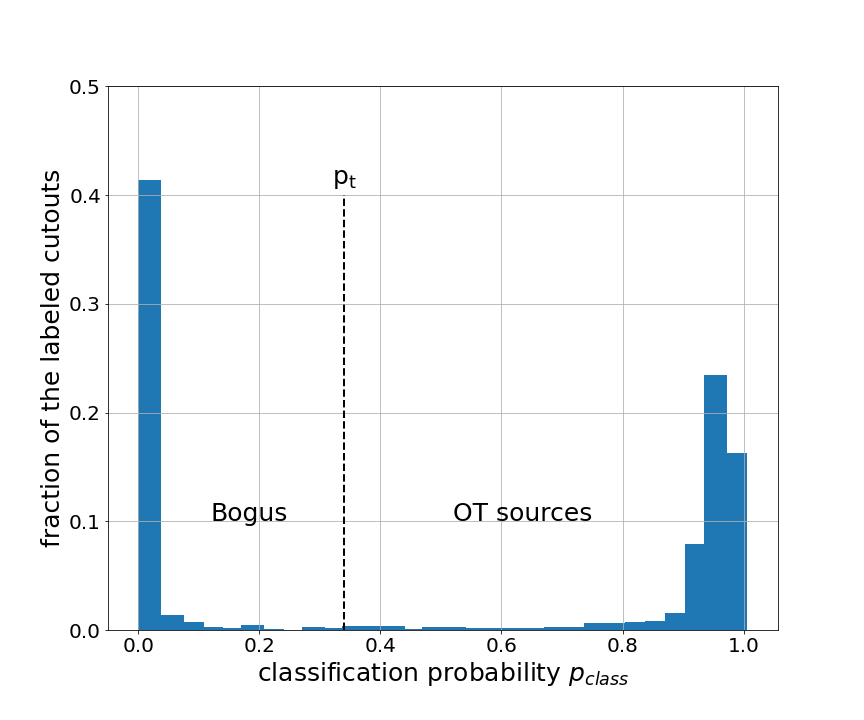}
    \end{minipage}
    \begin{minipage}{0.49\linewidth}
    \includegraphics[trim= 0 30 0 50, clip = true,width=1.0\textwidth]{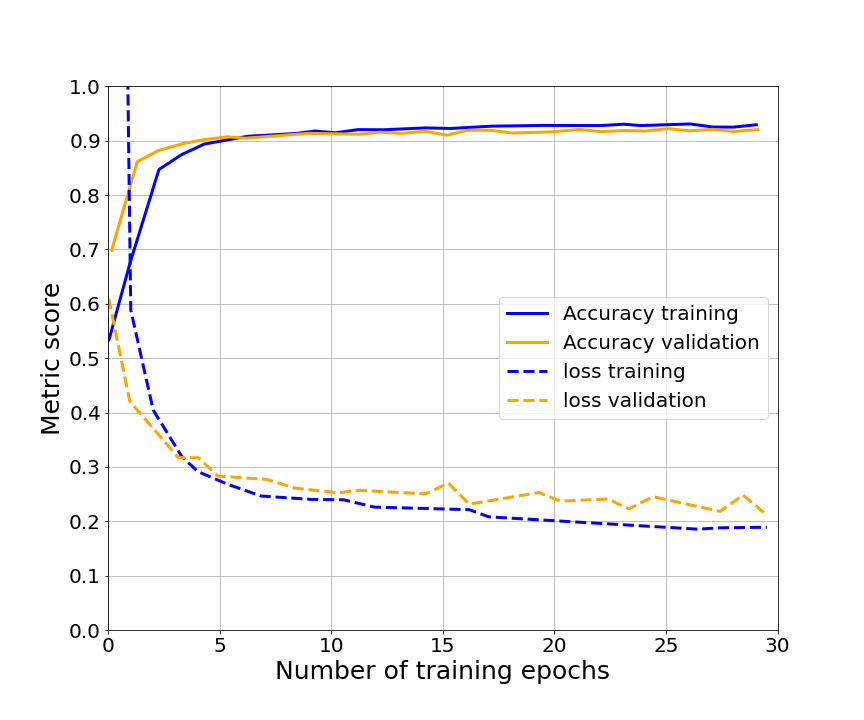}
    \end{minipage}
    \begin{minipage}{0.49\linewidth}
    \includegraphics[trim= 0 30 0 50, clip = true,width=1.0\textwidth]{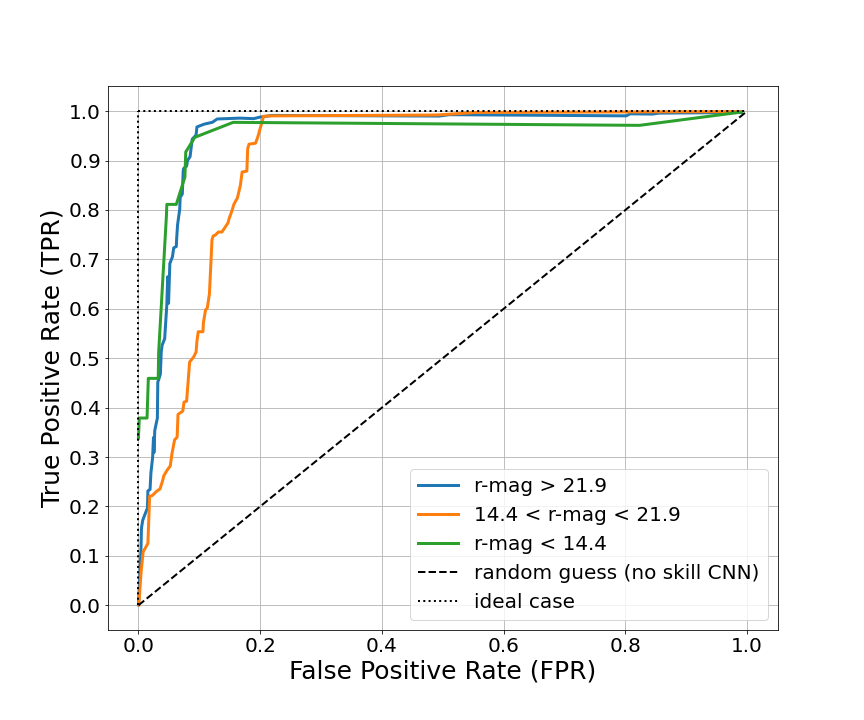}
    \end{minipage}
    \caption{Different visualization of the metrics used to evaluate the performance of the \otrain\ model trained on the TCA-{\sc gmadet} datacube. (Top left:) The evolution of the MCC and F1-score as a function of different values of $\mathrm{p_t}$. The value $\mathrm{p_t}=0.34$ (best threshold) maximizes both the MCC and F1-score. (Top right:) The classification probability distribution of the validation dataset at the last epoch of the training. (Bottom left:) Evolution of the Accuracy and loss metrics as a function of the number of training epochs. (Bottom right:) The ROC curve per bin of the OT candidate magnitudes in order to diagnose the behavior of the \otrain\ model for different source brightnesses.}
    \label{fig:TCA_res}
\end{figure*}

\end{appendix}

\end{document}